\documentclass[letterpaper]{JHEP3}

\usepackage{graphicx}
\usepackage{slashed}
\preprint{MIT-CTP-4024}

\title{Leptogenic Supersymmetry}

\author{Andrea De Simone\\
        Center for Theoretical Physics,
        Massachusetts Institute of Technology, Cambridge, MA 02139\\
        E-mail: \email{andreads@mit.edu}}
\author{JiJi Fan\\
        Department of Physics, Sloane Laboratory, Yale University, New Haven, CT 06520\\
        E-mail: \email{jiji.fan@yale.edu}}
\author{Veronica Sanz \footnote{On leave of absence from York University,  Ontario, Canada.}\\
         Physics Department, Boston University, 590 Commonwealth Ave, Boston,
        MA 02215\\
        E-mail: \email{vsanz@bu.edu}}
\author{Witold Skiba\\
        Department of Physics, Sloane Laboratory, Yale University, New Haven, CT 06520\\
        E-mail: \email{witold.skiba@yale.edu}}

\abstract{Leptogenic supersymmetry is a scenario characterized by cascade decays with copious lepton production. Leptogenic models have striking signatures that can be probed by the LHC even in the 10 TeV run with as little as $200 \textrm{ pb}^{-1}$ of data, provided the squark masses are about 1~TeV.\  Leptogenic supersymmetry spectrum arises in several well-motivated models and its signatures are long-lived sleptons, numerous isolated leptons, abundant Higgs production, rather energetic jets,  and no missing energy. The Higgs can be discovered in the $h\to b \bar{b}$ mode via the 4 leptons+4 jets channel because the leptons accompanying Higgs production suppress the background. The superparticle masses in leptogenic supersymmetry can be measured efficiently due to lack of missing energy and high lepton multiplicity. We estimate that $1 \,\textrm{ fb}^{-1}$ of integrated luminosity is sufficient to determine the light Higgs, neutralinos, charginos, slepton, sneutrino and squark masses in a 14~TeV run. }

\newcommand{\newc}{\newcommand}
\newc{ \GG         }{\tilde G}
\newc{ \Ni         }{ {\tilde N}_i }
\newc{ \Nj         }{ {\tilde N}_j }
\newc{ \NI         }{ {\tilde N}_1 }
\newc{ \NII        }{ {\tilde N}_2 }
\newc{ \NIII       }{ {\tilde N}_3 }
\newc{ \NIIII      }{ {\tilde N}_4 }
\newc{ \Ci         }{ {\tilde C}_i }
\newc{ \Cj         }{ {\tilde C}_j }
\newc{ \CI         }{ {\tilde C}_1 }
\newc{ \CII        }{ {\tilde C}_2 }
\newc{ \CIp        }{ {\tilde C}^{+}_1 }
\newc{ \CIpm       }{ {\tilde C}^{\pm}_1 }
\newc{ \CIm        }{ {\tilde C}^{-}_1 }
\newc{ \Cip        }{ {\tilde C}^{+}_i }
\newc{ \Cim        }{ {\tilde C}^{-}_i }
\newc{ \Cjp        }{ {\tilde C}^{+}_j }
\newc{ \Cjm        }{ {\tilde C}^{-}_j }
\newc{ \eL         }{ {\tilde e}_L }
\newc{ \eR         }{ {\tilde e}_R }
\newc{ \ser        }{ {\tilde e}_R }
\newc{ \smur       }{ {\tilde \mu}_R }
\newc{ \slr        }{ {\tilde \ell}_R }
\newc{ \sll       }{ {\tilde \ell}_L }
\newc{ \slep        }{ {\tilde \ell} }
\newc{ \veL        }{ {\tilde \nu} }
\newc{ \dL         }{ \tilde d_L }
\newc{ \dR         }{ \tilde d_R }
\newc{ \uL         }{ \tilde u_L }
\newc{ \uR         }{ \tilde u_R }
\newc{ \slepton    }{ \widetilde \ell }
\newc{ \ltilde     }{ {\tilde \ell} }
\newc{ \nutilde    }{ {\tilde \nu} }
\newc{ \snu        }{ { \tilde \nu} }
\newc{ \stau       }{ { \tilde \tau} }
\def\etmiss{\slashed{E}_T}
\def\ptmiss{\slashed{p}_{T}}

\newc{\winop       }{ \tilde W^+}

\newcommand{\beq}{\begin{eqnarray}}% can be used as {equation} or {eqnarray}
\newcommand{\eeq}{\end{eqnarray}}

\newcommand{\tev}{\,\mathrm{TeV}}
\newcommand{\kev}{\,\mathrm{keV}}
\newcommand{\gev}{\,\mathrm{GeV}}

\begin{document}

\section{Introduction}

\begin{figure}[ht]
\begin{center}
 \includegraphics[width=10cm]{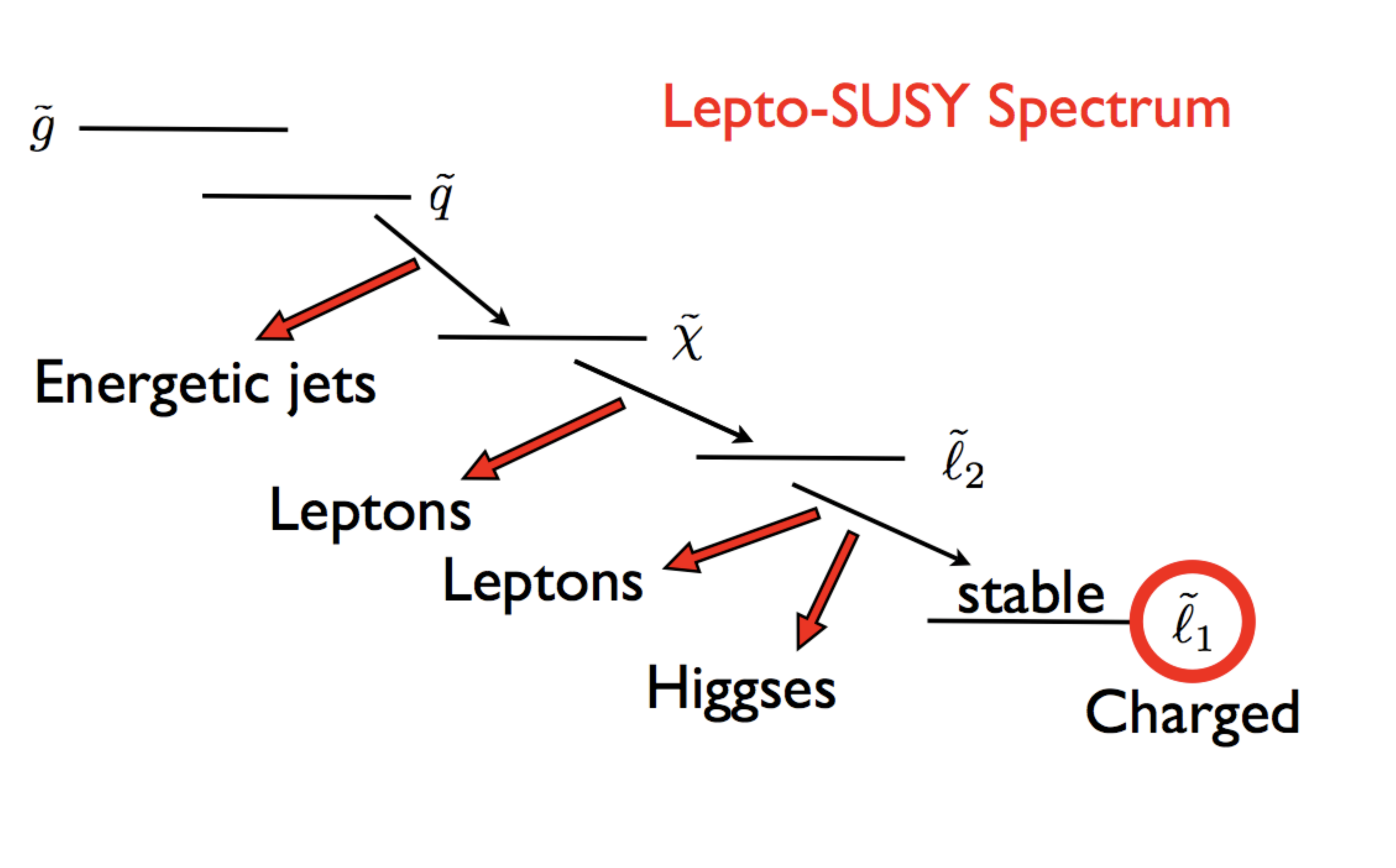}
\caption{Lepto-SUSY spectrum\label{spectrum} and typical decay channels.}
\end{center}
\end{figure}

What makes supersymmetry leptogenic is best explained pictorially in Fig.~\ref{spectrum}. The decay kinematics in  leptogenic supersymmetry (lepto-SUSY)  ensures that multiple leptons are produced in every decay chain. The production of new particles is dominated by QCD production of squarks and gluinos, which are assumed to be at the top of the mass spectrum. Colored particles decay into lighter charginos and neutralinos. The charginos and neutralinos
are heavier than the sleptons and therefore decay into leptons and sleptons. All sleptons decay into the lightest slepton which is the  next-to lightest supersymmetric particle (NLSP). The NLSP is collider stable and eventually decays into the gravitino. Because the gravitino LSP
is created outside detectors it is not shown in Fig.~\ref{spectrum}. 
  
This particular hierarchy of masses is responsible for collider signals with very little Standard Model (SM) backgrounds. For example, the leading production mechanisms, $\tilde{q} \, \tilde{q}$ and $\tilde{q} \, \tilde{\bar{q}}$, lead to at least two hard jets from the decays of the two squarks. The two jets are accompanied by two sleptons and a handful of leptons as every supersymmetric particle must decay into the slepton NLSP. Because of the large mass differences between the squarks and sleptons the NLSP sleptons are quite energetic and a large fraction of them have velocities larger than 0.95. Some of the fast sleptons would likely be misidentified as muons and we refer to the misidentified sleptons as leptons. The total number of observed leptons in an event varies depending on the details of the decay chain and how many of the leptons are neutrinos and taus, which are difficult to reconstruct.
One expects anywhere between two and eight observed leptons some of which would be misidentified sleptons, as well as up to two properly identified sleptons.  We analyze in detail events with 4, 5, and 6 lepton-like particles, as events with fewer identified leptons have non-negligible backgrounds and events with more leptons are rare.  

The decay chains that pass through the heavier sleptons lead to significant Higgs boson production. This happens most of the time for the heavy stau decays and a fraction of the smuon decays as long as the mass splitting between the sleptons is larger than the Higgs mass. The Higgs is then associated with a clean four-lepton signature and can be discovered in the $b \bar{b}$ channel. This is thrilling as for most other scenarios with a light Higgs, the $b \bar{b}$ would not be the first channel to be observed, and in fact that channel may never be observed. 

Assuming squark masses of about  1~TeV, we show that very significant excesses in every (4, 5, and 6) lepton channel can be obtained with only 200~pb$^{-1}$  of integrated luminosity in the 10~TeV run of the LHC. Some of the masses can be reconstructed with this little luminosity, while most masses can be determined using 1~fb$^{-1}$ of data collected at 14~TeV.\ 

Our approach is model independent as we parameterize the spectrum and do not make assumptions about its origin. However, lepto-SUSY spectrum does arise in several models. Low-scale gaugino mediation (L$\tilde{g}$M)~\cite{lgm} is a class of models with a parametric suppression of sfermion masses compared to the gauginos.  Other examples include gauge mediation (GMSB)~\cite{gmsb} with a large number of messengers as well as models with Dirac gaugino masses~\cite{Fox:2002bu}.  

The structure of this article is as follows. In the next section,  we present a sample spectrum and describe the most important slepton decay channels.  In Section~\ref{pheno}, we turn to LHC phenomenology. We discuss events with multiple leptons, analyze the most promising  Higgs discovery channel, and study spectrum reconstruction. Section \ref{sec:models} contains our parameterization of the soft mass terms. We also outline how the lepto-SUSY spectrum is featured in several mediation models. We  conclude and outline future directions in Section ~\ref{concl}.

%%%%%%%%%%%%%%%%%%%%%%%%%%%%%%%%%%%%%%%%%
%%%%%%%%%%%%%%%%%%%%%%%%%%%%%%%%%%%%%%%%%
%%%%%%%%%%%%%%%%%%%%%%%%%%%%%%%%%%%%%%%%%

\section{ Lepto-SUSY spectrum and decays}
\label{sec:setup}
The spectrum of lepto-SUSY is as depicted in Fig.~\ref{spectrum}, namely
\beq
m_{\tilde{g}}, m_{\tilde{q}}> m_{\tilde{\chi}^0},m_{\tilde{\chi}^{\pm}} > m_{\tilde{\ell}_L} > m_h, m_{\tilde{{\ell}}_R}\,. \nonumber
\eeq
The gluino can be either lighter or heavier than the squarks. Such ordering of masses emerges in several scenarios of supersymmetry breaking.
In Section~\ref{sec:models}, we propose a natural parameterization of the soft mass terms that is applicable to gauge mediated models. In our parameterization, the scalar soft masses are described by four dimensionless numbers and are proportional to the gaugino masses. The gaugino masses
are assumed to obey the unified relations so, together with $\tan \beta$ and ${\rm sign}\ \mu$, we have 7 parameters. The details are not
crucial for the phenomenological studies we are about to describe and are therefore postponed to Section~\ref{sec:models}.

We impose constraints on the spectrum that ensure correct electroweak symmetry breaking (EWSB), and that the slepton \cite{cernsearch} and Higgs masses are above the direct search limits. The Higgs mass is most sensitive to the the magnitude of the stop loop correction.  The bound $m_h>114 \gev$ implies a lower bound on $m_{\tilde q}$, which in the large $\tan\beta$ limit translates to $m_{\tilde q}>700$ GeV.  The lightest neutralino $\tilde \chi_1^0$ can be either Higgsino-like or Bino-like depending on the relative sizes of $\mu$ and $m_1$, where $m_i$ are the gaugino soft mass terms with $i=1,2,3$ corresponding to the SM gauge groups. If $\mu\gg m_1$, $\tilde \chi_1^0$ is mostly Bino-like while it is Higgsino-like for $\mu\ll m_1$.  
 
Using these results, we find the bounds on the mass ratios that are relevant to the production and decay channels we will consider: 
\beq
 3&>&m_{\tilde q}/m_2>1.1, \,
5.8>m_{\tilde q}/ m_1>2.2, \nonumber\\
12&>&m_1/ m_{\tilde \ell_R}>1,\quad 26>m_2/
m_{\tilde \ell_L}>1.9, \quad \mu/m_1>0.2. \label{eq:massrange}\eeq
When the constraints are taken into account, the gap between sleptons and gauginos can be rather large. 
The MSSM spectra are calculated using the SUSY-HIT program~\cite{susyhit} and a sample spectrum is shown
in Table~\ref{table:sample}. The soft masses and our parameters that yield the sample spectrum are described in Section~\ref{sec:models}.

\begin{table}[ht]
  \centering
  \begin{tabular}[c]{ | c | c |  c | c |}
  \hline
          $ m_{\tilde g}$ & $1938$ &  $ m_{\tilde{u}_L}$ & $949$ \\
         $ m_{\tilde{\chi}_1^\pm}$ & $291$ &  $ m_{\tilde{u}_R}$    & $920$ \\
         $ m_{\tilde{\chi}_2^\pm}$ & $676$ &  $ m_{\tilde{d}_L}$ & $952$  \\
         $ m_{\tilde{\chi}_4^0}$ & $676$  &  $ m_{\tilde{d}_R}$ & $919$ \\
         $ m_{\tilde{\chi}_3^0}$ & $ 353$ &   $ m_{\tilde{t}_1}$ & $920$ \\
  $ m_{\tilde{\chi}_2^0}$ & $302$  &  $ m_{\tilde{t}_2}$ & $962$  \\
    $m_{\tilde{\chi}_1^0}$  & $271$ &  $ m_{\tilde{\ell}_L}$ & $248$ \\
          $m_h$ &  $115$ & $ m_{\tilde{\ell}_R}$ & $108$  \\
          $m_{H^\pm}$ &  $387$ & $ m_{\tilde{\nu}}$ & $236$  \\
          $m_A$      & $379$ &  $ m_{\stau_1}$ & $106$  \\
          $m_{H_0}$  & $379$  &  $ m_{\stau_2}$ & $249$  \\
   \hline
  \end{tabular}
  \caption{A sample spectrum calculated with SUSY-HIT using input soft terms described in Section~\protect\ref{sec:models}: $\mu= 294$ GeV, $A = 0$, and $\tan \beta = 10$. All masses are in GeV.    \label{table:sample} }
\end{table}

In our study the trilinear A-terms are set to zero, so the three generations of squarks are nearly degenerate and there is little mixing between the left- and right-handed squarks. Thus all three generation squarks can be produced at a hadron collider. As a consequence, the Higgisinos with significant couplings to the third-generation squarks only could also be produced in cascade decays. Non-zero A-terms contribute the term $ A v \sin\beta$ to the off-diagonal entries in the squark mass matrix, where $v=174\gev$ is the EWSB vacuum expectation value. Assuming only the stop A-term, $A_t$, is large, a sizable mass splitting between  the up-squarks and stops of order $\delta m \sim 100 \gev$  would require $A_t v /(2m_{\tilde q})\sim100 \gev$ in the large $\tan\beta$ limit.  For our sample point, $m_{\tilde q} \sim 1 \tev$ which corresponds to $A_t \sim 1\tev$.  This is unnaturally large in models that lead to the lepto-SUSY spectrum, hence we neglect the effects of the A-terms.

In the sample spectrum in Table~\ref{table:sample}, the NLSP is $\tilde{\tau}_1$. 
The decay length of the NLSP, produced with energy $E$, in the laboratory frame is 
(see e.g. \cite{gmsb},~\cite{primer})
\beq  
L(\tilde{\tau}_1 \to \tau \tilde{G}) \simeq 1.7\,\sqrt{{E^2\over m_{\stau_1}^2}-1}
\left(\frac{100 \gev}{m_{\stau_1}}\right)^5\left(\frac{m_{3/2}}{10 \kev}\right)^2  \, 
\rm{km}\,,
\eeq
where $m_{3/2}$ is the gravitino mass. For the typical masses of our model, gravitinos lighter than 1 GeV are
cosmologically safe, i.e. they evade the constraints from overclosure of the
universe and Big Bang nucleosynthesis, provided that the reheating
temperature is lower than about $10^7$ GeV.

For simplicity, we will assume that $m_{3/2}\sim 10 \kev$.  
Then $\stau_1$ has a very long lifetime and exits detectors without decaying. The other right-handed sleptons, $\tilde{e}_R$ and $\tilde{\mu}_R$, will decay to $\tilde{\tau}_1$ through $\tilde{\ell}_R \to \ell_R\tau_R\tilde{\tau}_R$. Due to the small mass difference  $m_{\tilde{e}_R}- m_{\tilde{\tau}_1} -m_\tau=0.6$~GeV, the lepton and $\tau$ produced in the decays have insufficient energy to pass $p_T$ cuts. Larger mass splittings were considered in Ref.~\cite{Ambrosanio}. In our case, the decay lengths of $\tilde{e}_R$ and $\tilde{\mu}_R$ are too short to observe an independent track. Even if the decay lengths were long enough to observe tracks, the small mass difference means there will not be visible kinks. Therefore, the phenomenology will resemble that of the ``slepton co-NLSP scenario"
of the well-studied GMSB point $\rm{G2b}$~\cite{atlas-tdr, thomas, ellis}. Note, though, that the ordering of the light neutralinos/charginos and of the heavier sleptons masses is opposite in lepto-SUSY to that of the $\rm{G2b}$ point.
We now turn to the decays relevant for collider phenomenology.

%%%%%%%%%%%%%%%%%%%%%%%%%%%%%%%%%%%%%%%%%

\subsection{Decay chains}
In lepto-SUSY scenario, higgsinos and gauginos must decay to the lightest collider-stable SUSY particles, $\tilde{\ell}_1$=$\tilde{e}_R$, $\tilde{\mu}_1$, $\tilde{\tau}_1$, and hence produce leptons. The number of leptons depends on whether the neutralinos go through a short ($\tilde\chi \to \tilde{\ell}_1$) or through a long  ($\tilde\chi \to \tilde{\ell}_2 \to  \tilde{\ell}_1 $) decay chain. Additional leptons and Higgses are produced in the long decay chains. Phase space permitted, $\tilde{\tau}_2$ undergoes a two-body decay into $Z$ or $h$. Those channels are closed for $\tilde{e}_L$ so the three-body $\tilde{e}_L\to \ell \ell' \tilde{\ell}'_1$ decay dominates. For $\tilde{\mu}_2$, both the two- and three-body decays are open.

%%%%%%%%%%%%%%%%%%%%%%%%%%%%%%%%%%%%%%%%%
\subsubsection{Leptons galore}
\label{3body}

In lepto-SUSY, the heavier sleptons can only decay into two leptons and the collider-stable slepton. Same-sign leptons (SSL) are therefore produced at a higher rate than opposite-sign leptons (OSL), which is another remarkable feature of the lepto-SUSY phenomenology. 

In the following, we will present analytical results for the three-body decay widths of $\tilde{e}_L$ and $\tilde{\mu}_L$.  The effect of mixing will be discussed in Sec.~\ref{2body} as it is important for the staus and smuons. Here we use the notation of Refs.~\cite{Ambrosanio} and~\cite{HaberKane}.  

The number of SSL and OSL is given by 
 \beq  \Gamma_{\rm{SSL, OSL}} = \Gamma(\sll^- \rightarrow \ell_L^-
\ell_R^{\mp} \slr^{\pm}) \approx {m_{\sll} \over 512 \pi^3 } \sum_{i,j=1}^4
 c_{ij} I^{(1,2)}_{ij}, \label{fullwidth1}    
\eeq
 where we have adapted the results in Ref.~\cite{Ambrosanio}. The coefficients are
 \beq c_{ij} = g_1^{ 2} N_{j1}N_{i1}^*(g_1 N_{j1}+g_2N_{j2})(g_1 N^*_{i1}+g_2N^*_{i2}) ,\eeq
where $N_{ij}$ is the neutralino mixing matrix. 
 
 \begin{figure}[htbp]
\begin{center}
\includegraphics[scale=0.65]{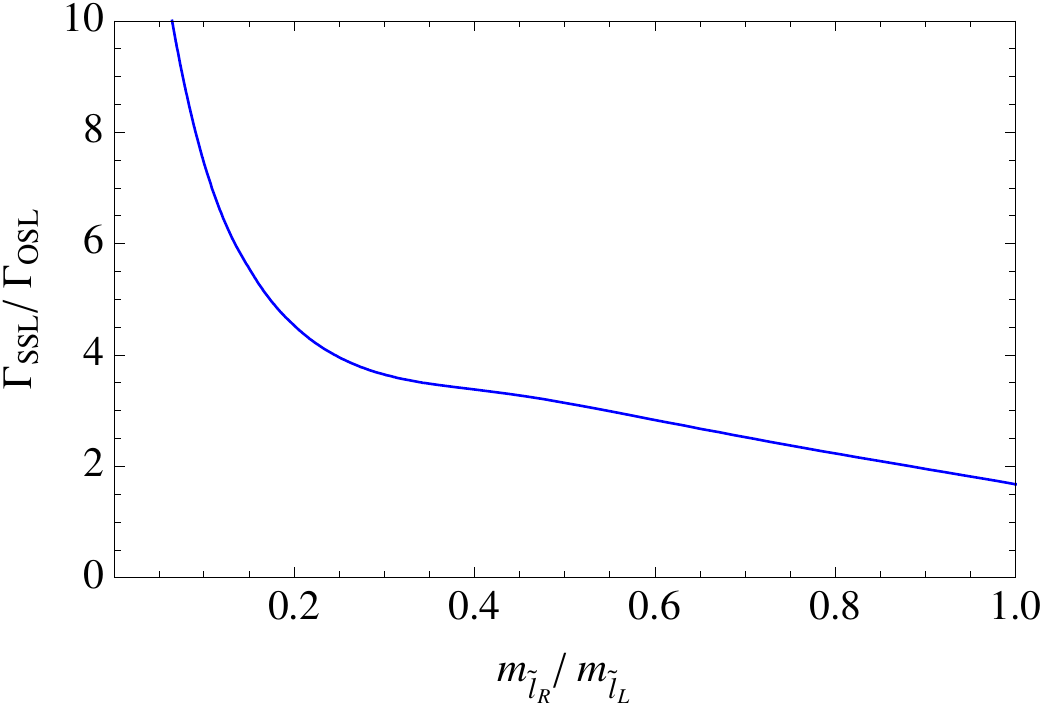}
\caption{ $\Gamma_{\rm{SSL}}/\Gamma_{\rm{OSL}}$ ratio as a function of
$m_{\tilde{\ell}_R}/m_{\tilde{\ell}_L}$.  The neutralino mass matrix corresponds to
our sample point. }
\label{higgsbbar}
\end{center}
\end{figure}
 
 These three-body decays are typically mediated by the Bino and, for Majorana neutralinos,  are suppressed by its mass with $\Gamma_{OSL} \sim  {m_{B}^{-4}}$ while $\Gamma_{SSL} \sim  {m_{B}^{-2}}$. The Bino mediated decays are partially responsible for the SSL  excess over OSL in lepto-SUSY. 
For our sample point, where $m_{\tilde{\ell}_R}/m_{\tilde{\ell}_L}\simeq 0.44$, the SSL final state branching ratio is about $3.6$ times that of OSL for both selectron and smuon three-body decays.

%%%%%%%%%%%%%%%%%%%%%%%%%%%%%%%%%%%%%%%%%

\subsubsection{Higgses galore}
\label{2body}

Even though the decay of the heavier sleptons into the Higgs boson and the light
sleptons is suppressed by the small Yukawa couplings such decay is competitive with the three-body
decay channels for the $\tilde{\mu}_2$, and it completely dominates for the $\tilde{\tau}_2$.  Below, we derive the  $\tilde{\tau}_2$  two-body decay
widths into the Higgs and the Z bosons. Analogous results apply to the smuons.  The mass eigenstates are defined as:
\beq
\pmatrix{\stau_1\cr \stau_2} = \pmatrix{\cos\theta_\stau &
\sin\theta_\stau \cr
         -\sin\theta_\stau & \cos\theta_\stau}
\pmatrix{\stau_R\cr \stau_L}, \nonumber \eeq where $\theta_\stau$
is the mixing angle, $0\leq\theta_\stau<\pi$, and $m_{\stau_1}<m_{\stau_2}$. To the leading order,
\beq 
 \sin \theta_\stau \approx \frac{\mu\,  m_\tau \tan\beta}{m_{\stau_2}^2-m_{\stau_1}^2} \ \ \ {\rm and} \ \ \ \cos \theta_\stau \approx 1.
\eeq 
The two-body decay widths are then
\begin{eqnarray}
\Gamma(\stau_2 \rightarrow \stau_1+h) &=& \frac{\mu^2y_\tau^2\cos^2\alpha\cos^2(2\theta_\stau)}{16\pi m_{\stau_2}} f_1(r_{\stau_1},r_h) \, , \nonumber\\
\Gamma(\stau_2 \rightarrow \stau_1+Z) &=&
\frac{g_2^2\sin^2(2\theta_\stau)m_{\stau_2}^3}{128\pi\cos^2\theta_W
m_Z^2} f_2(r_{\stau_1},r_Z) \, .
\end{eqnarray}
 where $\theta_W$ is the weak mixing angle, while the mass
ratios are $r_{\stau_1}=m_{\stau_1}/m_{\stau_2}$ and
$r_{h/Z}=m_{h/Z}/m_{\stau_2}$. The dimensionless functions
$f_1,f_2$ are defined as
 \begin{eqnarray}
f_1(r_{\stau_1},r_h)&=&\sqrt{\frac{1}{4}\left(1-r_{\stau_1}^2+r_h^2\right)^2-r_h^2}\, ,\\
f_2(r_{\stau_1},r_Z)&=&\left((1-r_{\stau_1}^2)^2-2
r_Z^2-2r_{\stau_1}^2 r_Z^2+r_Z^4\right)f_1(r_{\stau_1},r_Z)\, .
\end{eqnarray}
In the decoupling limit, where $c_{\alpha} \sim s_{\beta}$, one
finds that $\tan \beta$ drops out of the ratio of the two decay
widths:
\begin{equation}
\label{ } \frac{\Gamma(\stau_2 \rightarrow \stau_1+Z)
}{\Gamma(\stau_2 \rightarrow \stau_1+h) } =\frac{
\left((1-r_{\stau_1}^2)^2-2 r_Z^2-2r_{\stau_1}^2
r_Z^2+r_Z^4\right)}{(1-r_{\stau_1}^2)^2}\frac{f_1(r_{\stau_1},r_Z)}{f_1(r_{\stau_1},r_h)}\, .
\end{equation}
For the sample spectrum we chose, $\tilde{\tau}_1$ decays into
Higgses $53\%$ of the time while for $\tilde{\mu}_1$, including three-body decays, the branching ratio to Higgses is $44 \%$.

%%%%%%%%%%%%%%%%%%%%%%%%%%%%%%%%%%%%%%%%%
%%%%%%%%%%%%%%%%%%%%%%%%%%%%%%%%%%%%%%%%%
%%%%%%%%%%%%%%%%%%%%%%%%%%%%%%%%%%%%%%%%%

\section{LHC phenomenology}
\label{pheno}

Our analysis of collider signatures was performed using the following software tools.  We generated the events at the parton level with the Monte Carlo generator MadGraph \cite{madgraph}. We used a modified version of  BRIDGE \cite{bridge} for particle decays to account for the $\tilde{\mu}_L-\tilde{\mu}_R$ mixing. Then, we passed the events through PYTHIA \cite{pythia} to include showering-hadronization effects. Finally, we estimated the detector effects following ATLFAST \cite{atlfast} approach. For example, to include energy resolution effects we modified the ATLFAST subroutines for the jet and electron energy smearing. The analysis of the events is performed with basic parton level $\eta$ and $p_T$ cuts on jets, leptons, and sleptons. We also imposed jet and lepton isolation cuts.

\begin{figure}[t!]
\begin{center}
\includegraphics[width=1.5in,angle=90]{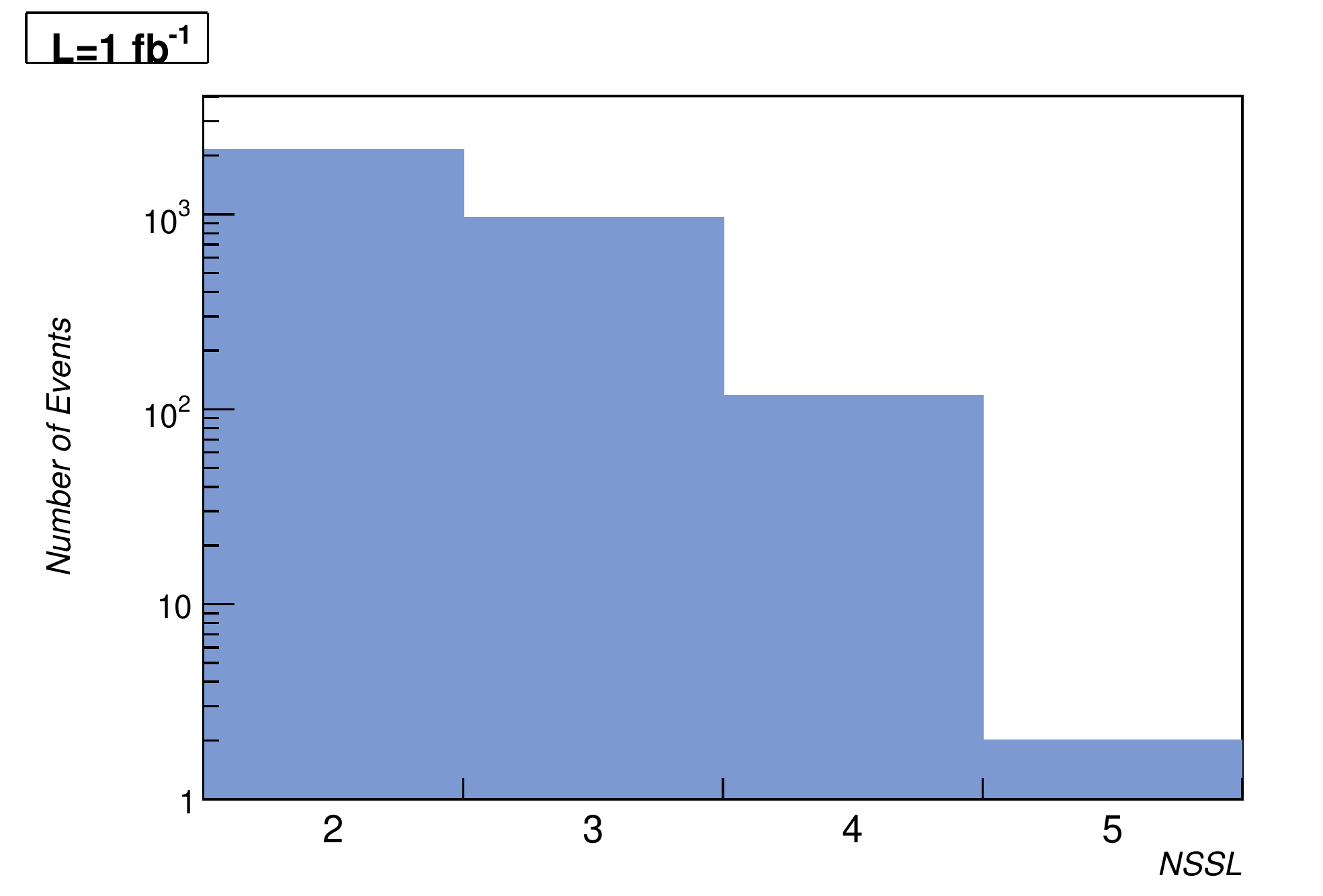}
\includegraphics[width=1.5in,angle=90]{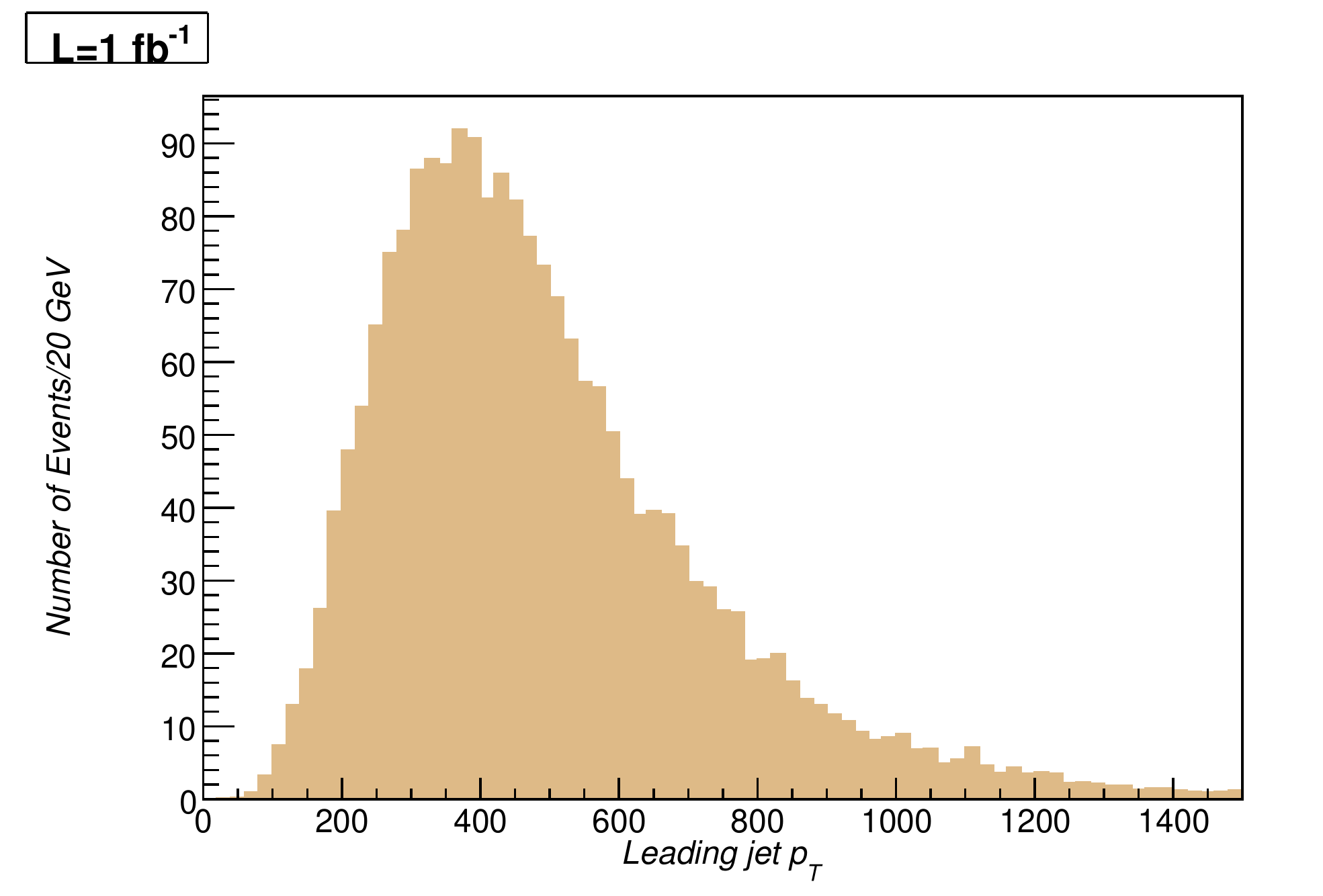}
\caption{Number of same-sign leptons and leading jet $p_T$ distributions for  $\sqrt{s} = 14$ TeV and ${\cal L} = 1 \, \rm{fb}^{-1}$.}
\label{ptjet}
\end{center}
\end{figure}

We focus on events characterized by 
\begin{enumerate}
  \item Large cross section production from squark pair production.
  \item Two hard jets (see Fig.~\ref{ptjet} for the $p_T$ distribution of jets from the squark decays).
  \item  At least four lepton-like particles (leptons or stable sleptons).
 \end{enumerate}

We concentrated on these signals because they are practically background free. There are no sources of, real or fakes, four-lepton events with at least two SSL and two hard jets with the cross sections in the fb range (see discussion in Section \ref{sec4lep}). Our goal is twofold. First, we want to show that statistically significant excesses of events can be observed
with little integrated luminosity. Second, we want to demonstrate that mass reconstruction of several states is possible. In order to do that, we 
examine the events with four, five, or six lepton-like particles. Such events probe different decay cascades and are therefore sensitive to different intermediate states. 

%%%%%%%%%%%%%%%%%%%%%%%%%%%%%%%%%%%%%%%%%

\subsection{Sleptons or muons?}

Long-lived sleptons are a promising feature of some of GMSB benchmark points~\cite{atlas-tdr,slep,cms-tdr}. In lepto-SUSY, the NLSP is a long-lived stau. Mass splittings between selectrons, smuons, and staus are rather small, so one cannot observe $\tilde{e}_R$ and  $\tilde{\mu}_R$ decays. As we previously mentioned, the phenomenology will resemble the so-called ``slepton co-NLSP scenario.''    A novel feature of lepto-SUSY is that pairs of sleptons are always accompanied by leptons. We presume the existing studies on long-lived sleptons in GMSB should be modified to account for high lepton multiplicity in this scenario, but we expect the essential features will remain unchanged. 

 \begin{figure}[t!]
\begin{center}
\includegraphics[width=1.5in,angle=90]{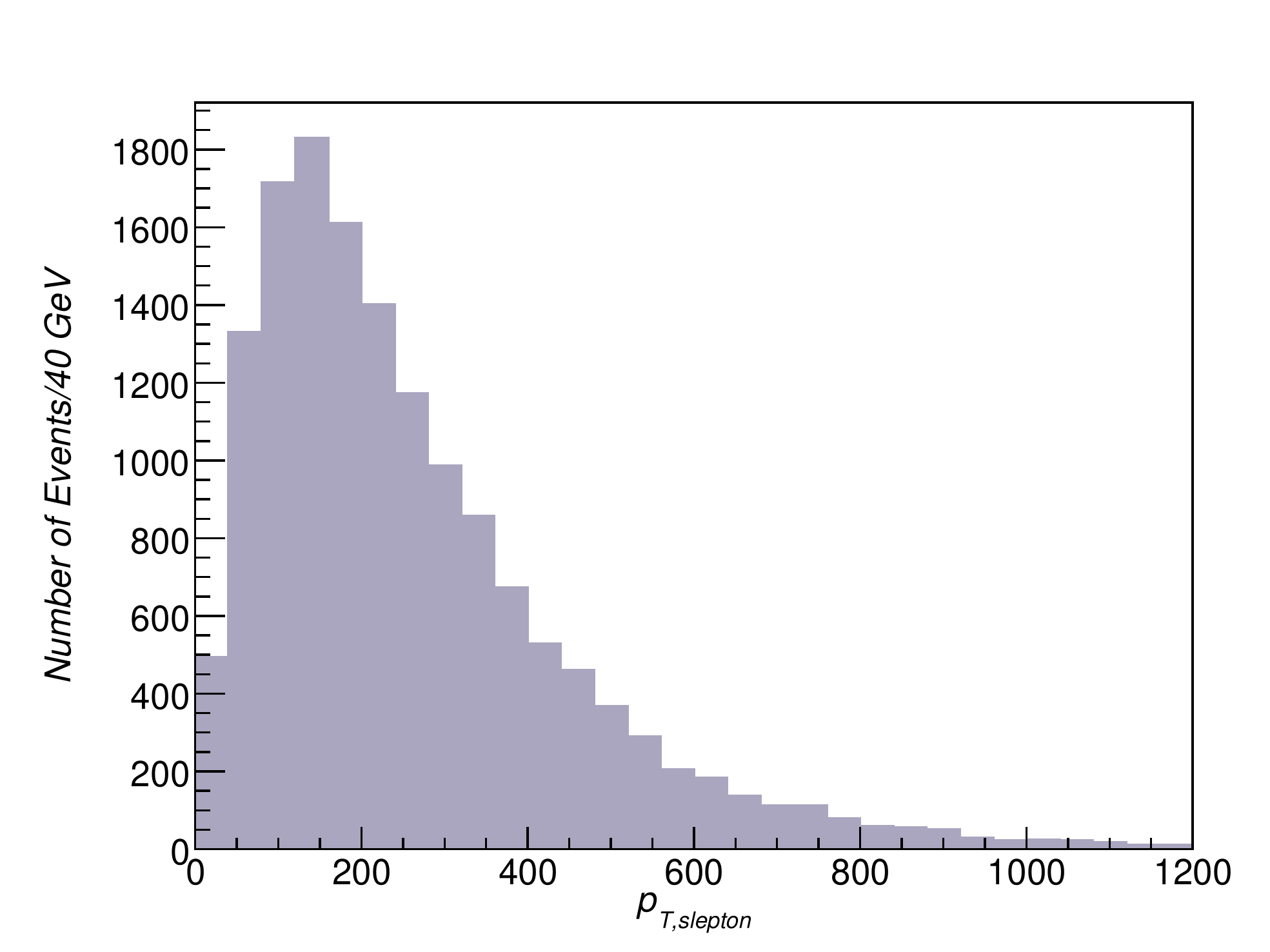}
\includegraphics[width=1.5in,angle=90]{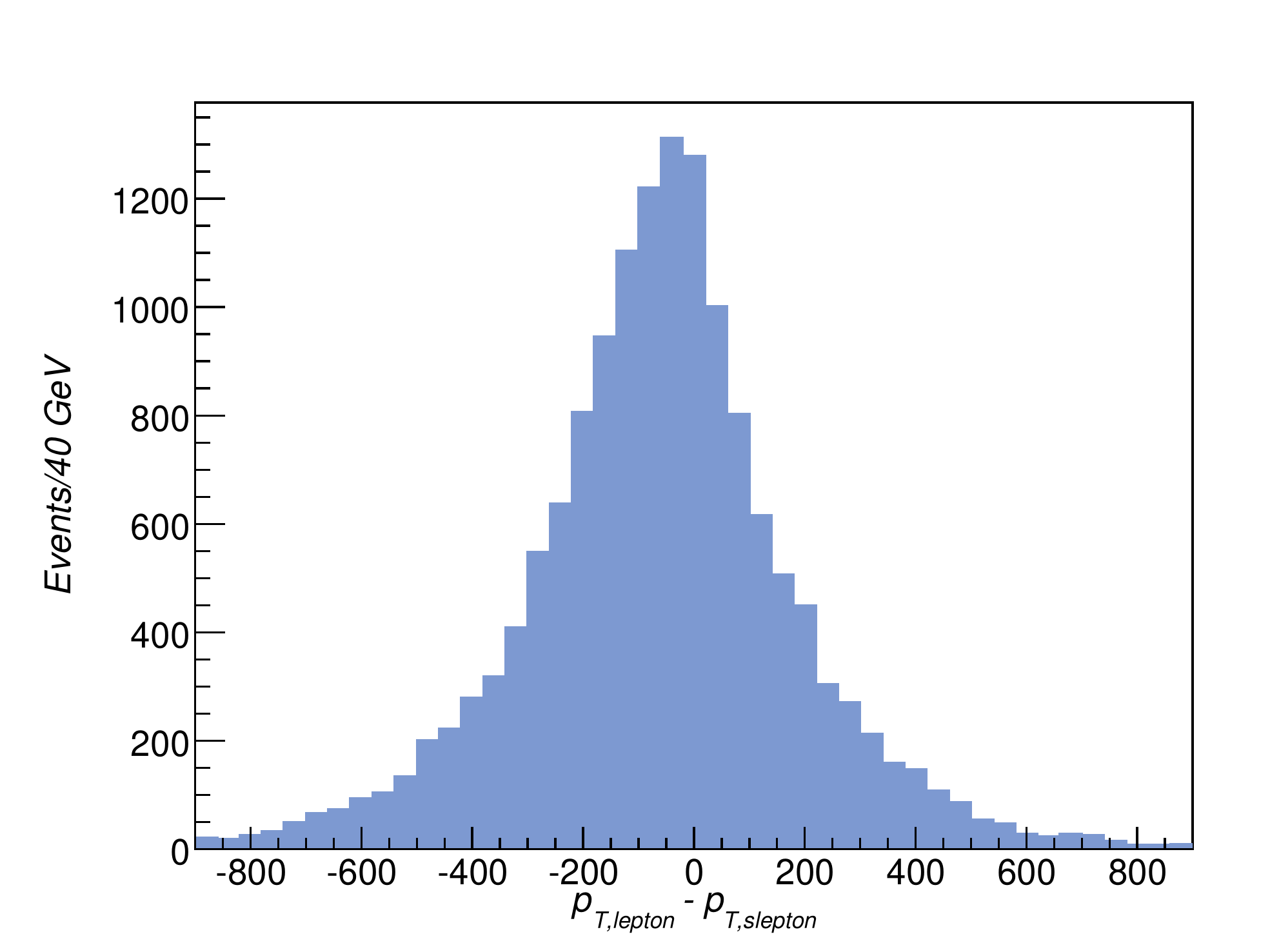}
\includegraphics[width=1.5in,angle=90]{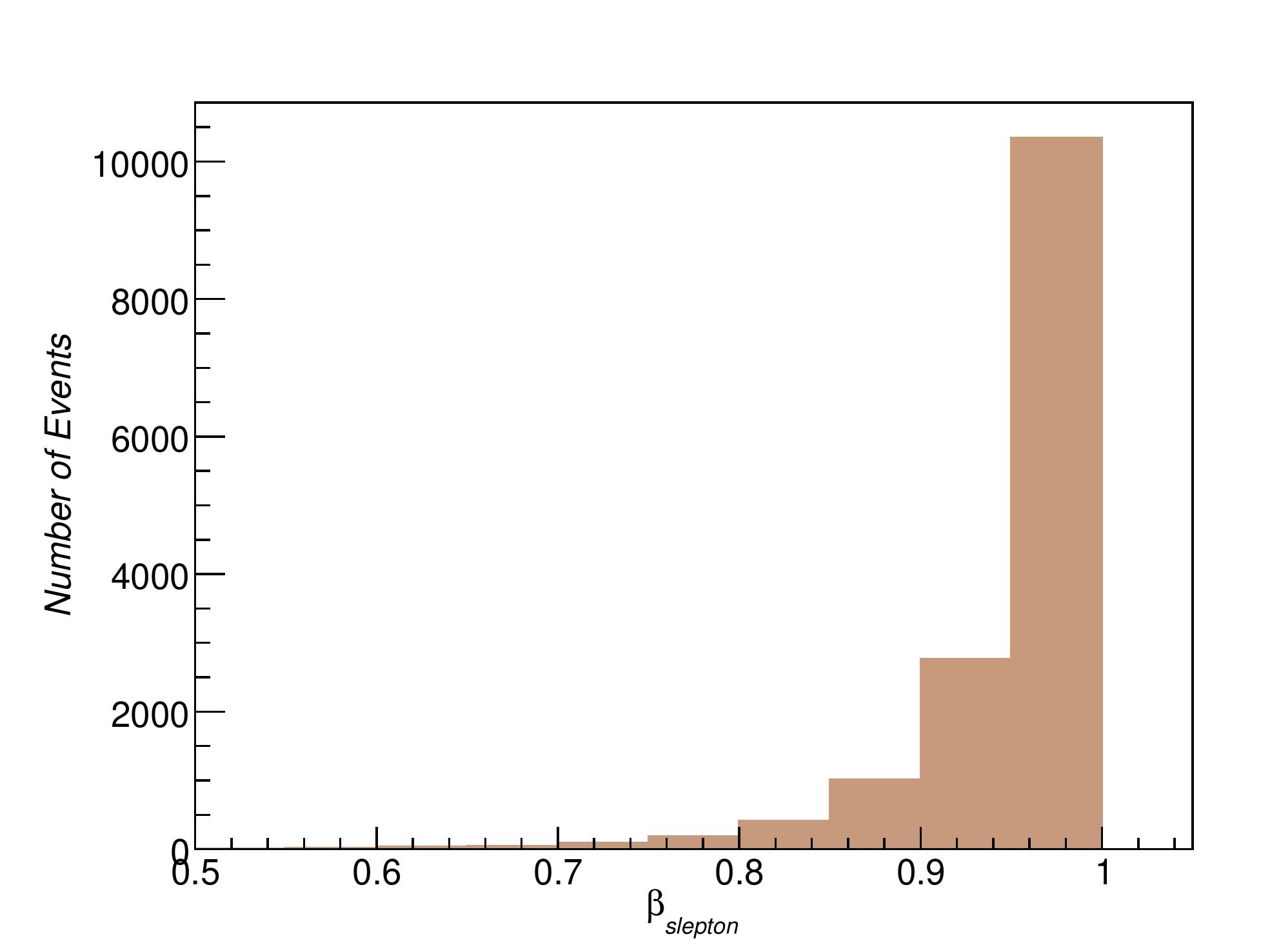}
\caption{Slepton $p_T$, lepton minus slepton $p_T$,  and slepton velocity distributions assuming  $\sqrt{s} = 14$ TeV and ${\cal L} = 10\, \rm{fb}^{-1}$.}
\label{betapt}
\end{center}
\end{figure}

Heavy, collider-stable particles appear as muons with a delayed arrival at the muon chambers. The dominant SM backgrounds are muons from the $b$ and $W$ decays. Cuts on the muon $p_T$ and isolation requirements greatly reduce these backgrounds. To reduce the background from $b$-decays, one applies cuts on the tranverse momentum of typically $p_T>$50 GeV~\cite{Allanach}. On the other hand, in our case the slepton is detected in association with leptons. This will reduce b-decay backgrounds and a relaxed $p_T$ cut is likely to suffice. Top decays have a higher $p_T$, but at the same time they have a smaller production cross section. The left plot in Fig.~\ref{betapt} shows the efficiency is high with a $p_T>$~50 GeV cut.

For stable particles one can infer their mass by  measuring the momentum and velocity. The momentum is measured with the tracker and the muon spectrometer.  The velocity can be measured using two techniques: time of flight (TOF) from the muon system or using the ionization information from the silicon tracker \cite{atlas-tdr,cms-tdr}. Cosmic muons are the main background for the TOF techique. To reduce this background, TOF measurement is often correlated with an independent measurement of ionization in the tracker. At CMS, this approach is suitable for velocities $0.6<\beta<0.8$~\cite{cms-tdr}.  Further refinements could be used to extend the range of $\beta$ to about $0.9$~\cite{tulika}.  At ATLAS, TOF technique may be applicable up to  $\beta$ of  around $0.95$~\cite{atlas-tdr}.

Sleptons in lepto-SUSY are decay products of heavy squarks. Hence, the slepton $\beta$ distribution is peaked towards large values. Indeed, most of our sleptons have velocities  $\beta>0.95$ and would therefore be likely misidentified as muons. To retain as much of the signal as possible, we treat the fastest sleptons with $\beta>0.95$ as muons while we assume that the slower sleptons are properly identified. We also use the muon smearing, isolation, efficiencies and charge misidentification parameters in the slepton analysis. Nevertheless, $30 \%$ ($57 \%$) of our sample contains at least one slepton with $\beta<0.8$ ($<0.9$) and an early slepton mass measurement is possible. Only 3$\%$ of the events contain 2 sleptons with velocities  less than $0.8$. In the dominant decay channels, as in $2\, \ell+ 2 \, \tilde{\ell} _R+ 2\, j$ channel, the total cross section of events with at least one slepton with $\beta<0.9$ is approximately 150 fb.

Another challenge associated with slow-moving long-lived particles is the correct assignment of bunch crossing. The efficiency for assigning the correct bunch crossing for slow moving particles decreases steeply with $\beta$. For $\beta \sim 0.8-1.0$, the efficiency is in the $80-100 \%$ range, whereas for $\beta\sim0.6$ it is only $15\%$ \cite{atlas-tdr}. A cut on the slepton $\beta>0.8$ eliminates very little signal, as illustrated by the right plot in Fig.~\ref{betapt}.
 
%%%%%%%%%%%%%%%%%%%%%%%%%%%%%%
\subsection{Production mechanism of lepto-SUSY}
\label{prod}

\begin{figure}[h!]
\begin{center}
\includegraphics[scale=1.0]{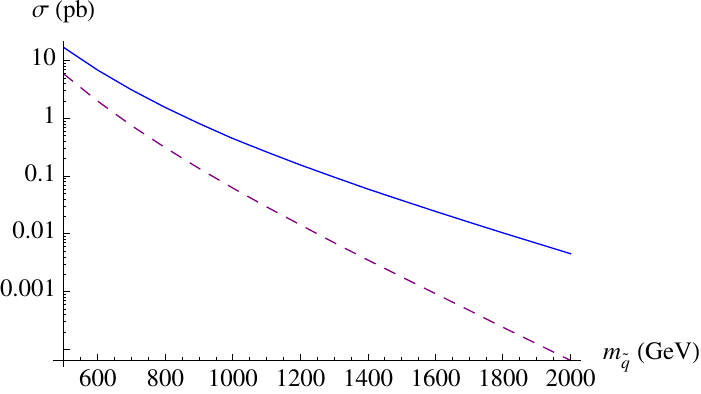}
\caption{The production cross sections $\sigma (pp\rightarrow\tilde{q}\bar{\tilde{q}}+\tilde{q}\tilde{q}+\bar{\tilde{q}} \, \bar{\tilde{q}} )$  (solid blue line) and  $\sigma (pp\rightarrow\tilde{q}\tilde{g})$ (dashed purple) at $\sqrt{s} = 14$ TeV for $m_{\tilde{g}}/m_{\tilde{q}}=2$  using Prospino~\protect\cite{prospino} at NLO.}
\label{prospino}
\end{center}
\end{figure}

The main production mechanism for lepto-SUSY events is pair production of squarks
\beq
pp\rightarrow \tilde{q} \, \bar{\tilde{q}} \ ,  \tilde{q} \, \tilde{q}, \, \bar{\tilde{q}} \, \bar{\tilde{q}}  \nonumber
\eeq
while the  $pp\rightarrow \tilde{g} \, \tilde{q}$ channel is suppressed when gluinos are heavier than squarks, see Fig.~\ref{prospino}. The cross sections for 
$pp\rightarrow \tilde{q} \, \bar{\tilde{q}} $ and $pp\rightarrow \tilde{q} \, \tilde{q} $ are of the same order for sparticle masses typical in lepto-SUSY. In the study we present in the next sections, production cross sections were computed within MG at the leading order and they are compatible with PROSPINO with a k-factor of ${\cal O}(1)$.

%%%%%%%%%%%%%%%%%%%%%%%%%%%%%%%%%%%%%%%%%

\subsection{Four-lepton channels}
\label{sec4lep}
\begin{figure}[h]
\begin{center}
 \includegraphics[scale=0.6]{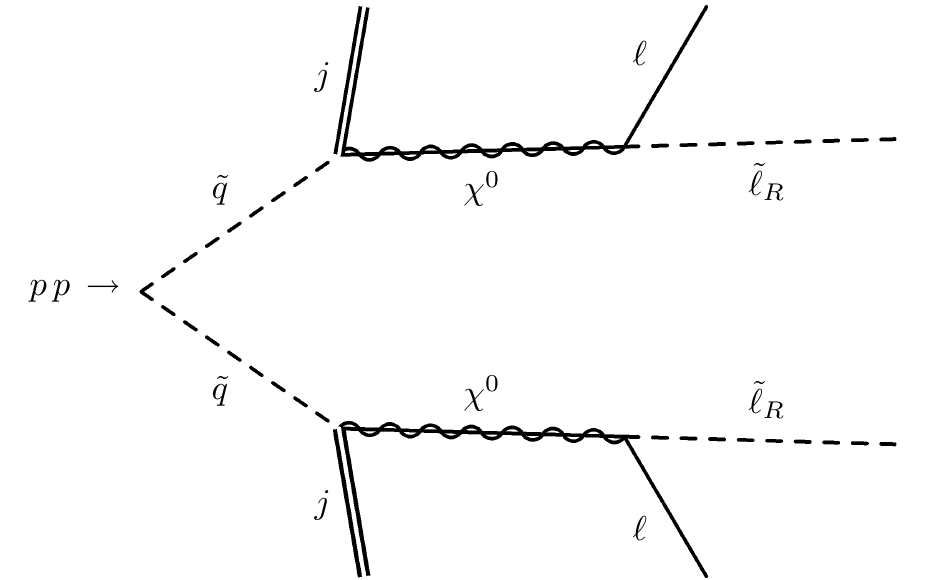}
   \includegraphics[scale=0.25,angle=90]{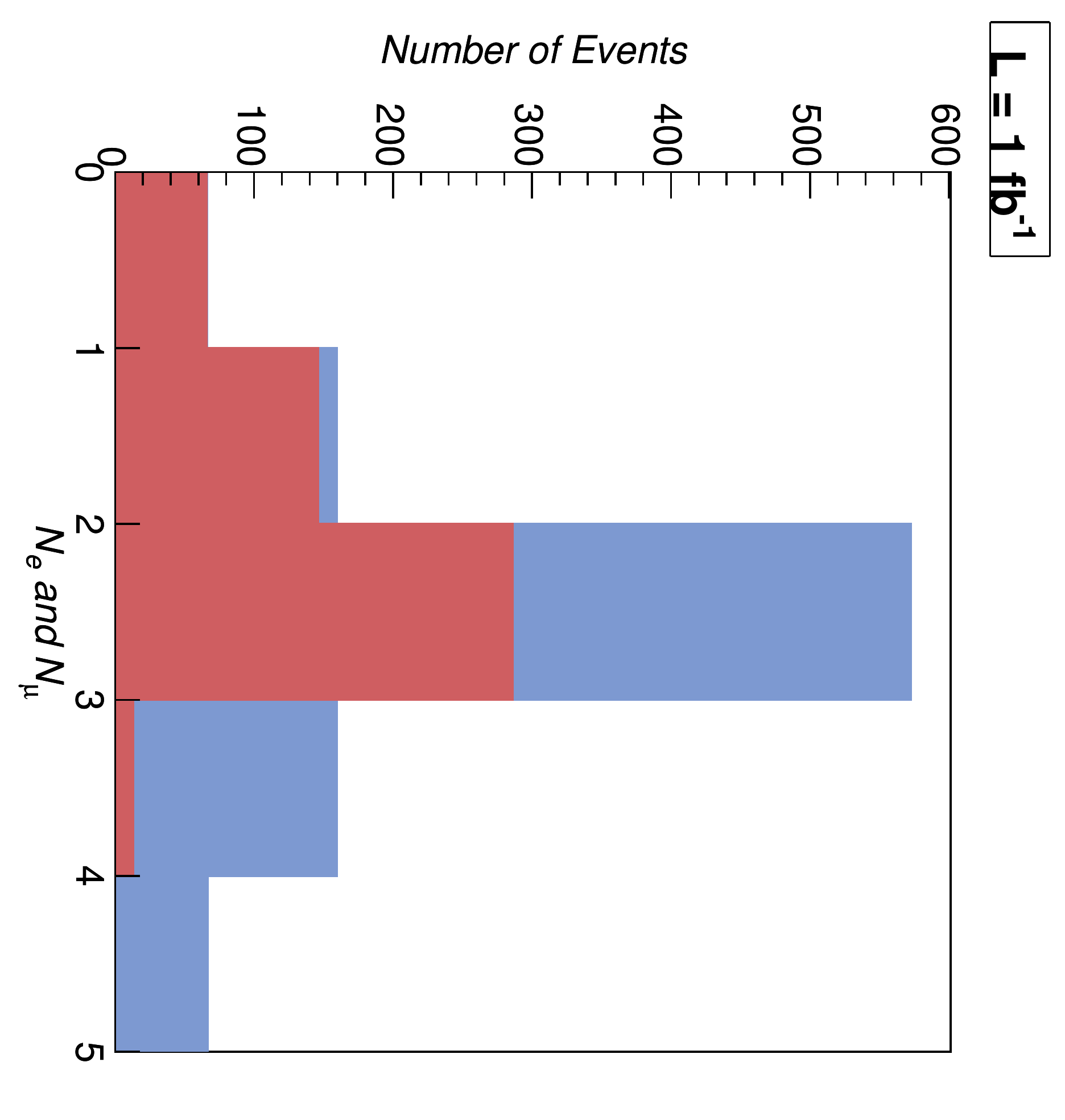}
   \hspace{0.5cm}
\caption{Left: Four-lepton channels. Right: Electron (lower-red) and muon (upper-blue) composition of the four leptons.}
\label{2lep-feyn} 
\end{center}
\end{figure}

Events with four lepton-like objects are best suited for reconstruction of masses of the initial colored and secondary color-neutral  particles---squarks and neutralinos. We refer to electrons, muons, and sleptons as lepton-like particles no matter if the sleptons are correctly identified or not. The four-lepton events  arise from short decay chains in which both neutralinos decay directly to heavy stable sleptons $\tilde{\ell}_R$ and leptons (that is  $e$ and $\mu$ only)
as depicted in Fig.~\ref{2lep-feyn}. 
Fig.~\ref{2lep-feyn} also shows the muon and electron composition of the four-leptons events where muons include sleptons. 
Other diagrams could potentially \textit{contaminate} this signal. For example, production of $\tilde{\chi}^0_4$ and $\tilde{\chi}^0_2$ neutralinos decaying into
\beq
\tilde{\chi}^0_4  \to 2 \, \ell \, \tau \, \tilde{\tau}_R\, , \nonumber \\
\tilde{\chi}^0_2 \to \tau \, h \,   \tilde{\tau}_R \, . \nonumber 
\eeq
Since $\tilde{\chi}^0_4$ is wino-like, it does not decay to $\tilde{\ell}_R$ directly but instead decays to $\tilde{\ell}_L$ and then to $\tilde{\ell}_R$. Those types of events occur at a lower rate than the signal we are interested in,  therefore they do not pose a problem. 

Events depicted in Fig.~\ref{2lep-feyn} do not have any missing energy hence cutting on $\etmiss$ would not affect our signal. Moreover, accurate missing energy calibration will be difficult in the early running, especially when hard jets are present in the event. Fortunately, we found that the selection of events in terms of the number of leptons is robust under different $\etmiss$ cuts. As a result, we do not impose a $\etmiss$ cut in our analysis of lepton channels. We will come back to the  $\etmiss$ cut in the Higgs search discussion in Section~\ref{sec:Higgs}.

We select four-lepton events using following criteria 
\beq
n_{l} &= &4\ (\rm{including\ sleptons}) \textrm{ with } |\eta|< 2.5, \textrm{ }  p_T> 10 \textrm{ GeV}  \nonumber \\
 &  &\textrm{ and parton level isolation cuts } \Delta R_{\ell\ell}>0.4, \, \Delta R_{\ell j}>0.4, \nonumber \\
n_{j} &\geqslant& 2 \textrm{ with } |\eta|< 2.5, \textrm{ }  p_T> 15  \textrm{ GeV} \nonumber \\
& & \textrm{ and  post-PYTHIA isolation cuts } \Delta R_{jj}>0.4.
\label{mincuts}
\eeq
With those cuts, the total cross section is 220 (690) fb  for $\sqrt{s}=10\ (14) \tev$ which corresponds to about 45 (140) events in 200 pb$^{-1}$ of collected data.  Notice very efficient hard cuts on the leading jets $p_T$ can be applied, see Fig.~\ref{ptjet}.

There are sources of SM backgrounds for events with 4 leptons+ 4 jets. After applying the cuts described in Eq.~(\ref{mincuts}), we estimated all SM backgrounds are below the fb level. We generated the most important backgrounds with ALPGEN~\cite{Mangano:2002ea}:  $t \bar{t}$+jets, $W$+jets, $W$$Z$+jets and QCD jets.  Other SM backgrounds like $Z$$Z$+jets are tiny, even before we ask for a hard cut on the jets-- see Fig. \ref{ptjet} in Ref.~\cite{Mangano:2002ea}. 

More relevant backgrounds come from QCD with jets faking leptons. We used a conservative estimate of $10^{-4}$ probability of that to happen \cite{atlas-tdr}.
Although QCD jets have a cross section of about  $10^8$ pb,  requiring four fake leptons cuts down their rate to several orders of magnitude below $1$ fb. Similar fake rate in backgrounds like $W$+jets or $W$$Z$+jets leads to a cross section significantly below the signal.

The $t \bar{t}$+jets background has an initial cross section of about 1 nb. We estimated the $b$-jets produce an isolated lepton in about 5$\times10^{-3}$. For $W$ decaying leptonically and the $b$'s producing isolated leptons, we estimate a cross section of about 1fb, which can be further reduced by applying a cut on the jets $p_T$.

Depending on whether sleptons are identified or not, we used different strategies to reconstruct the neutralinos participating in the diagram in Fig.~\ref{2lep-feyn}. In the absence of SM background, combinatorial background is the main obstacle to the reconstruction. 

Decay products of the squark decays are rather energetic---see Fig.~\ref{betapt}---and tend to cluster. Objects coming from a common decay tend to be near each other in the $\Delta$R parameter space. We used this information to reduce combinatorics. For the events with identified sleptons ($\beta \leq 0.95$),  we paired a slepton with a nearby lepton through $\Delta$R$_{\ell \tilde{\ell}}$ selection.  We followed the same procedure to pair the slepton-lepton pair with the nearest jet. For the events with sleptons misidentified as muon, we formed dilepton invariant masses and selected dilepton pairs with opposite charge and smaller $\Delta$R$_{\ell \ell}$. The $\Delta$R discrimination is very powerful. Simply taking the average of pairing each lepton with any opposite sign lepton would lead to no distinguishable features in the invariant mass distribution.

\begin{figure}[t]
\begin{center}
 \includegraphics[scale=0.3]{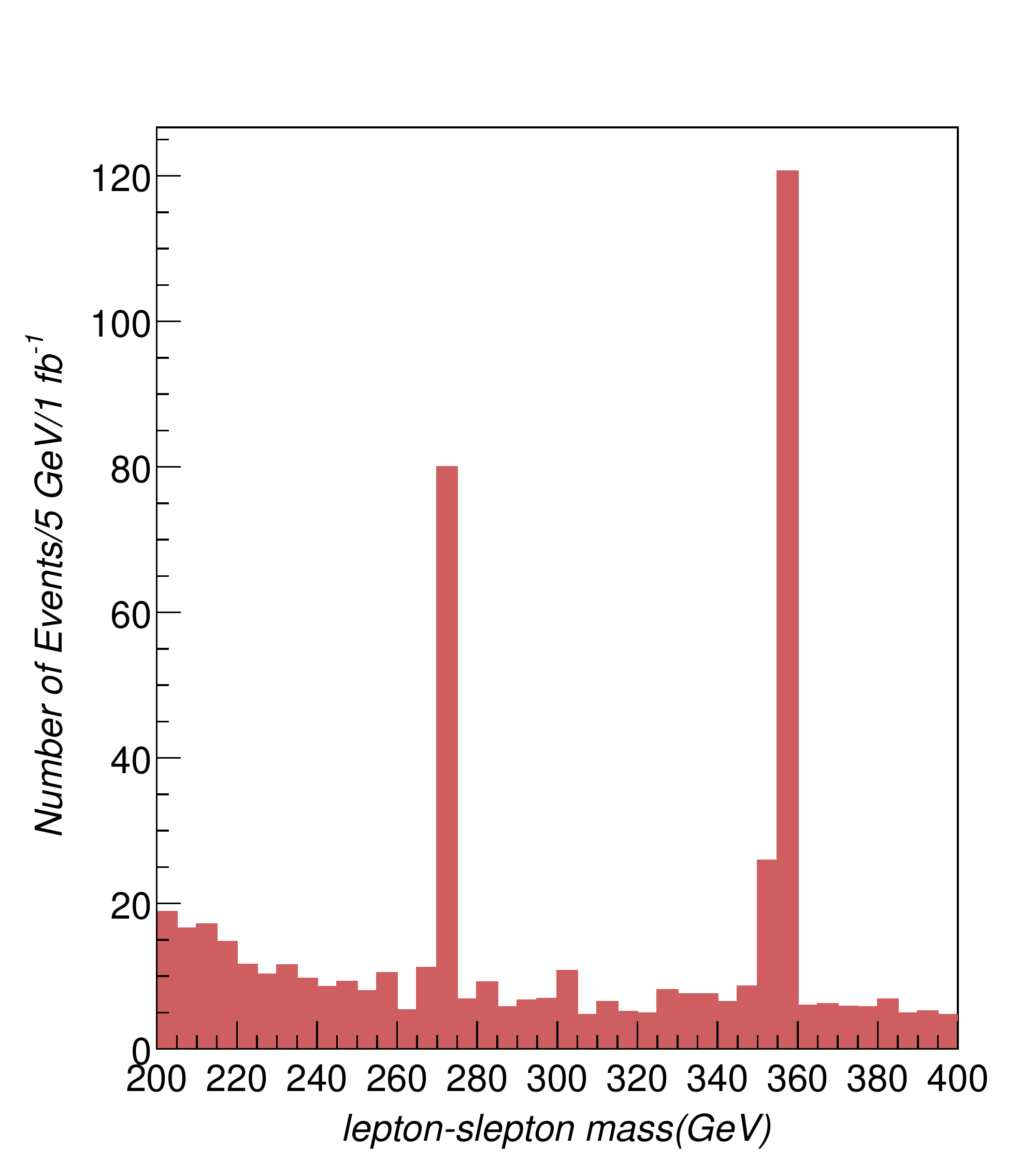}
 \includegraphics[scale=0.3]{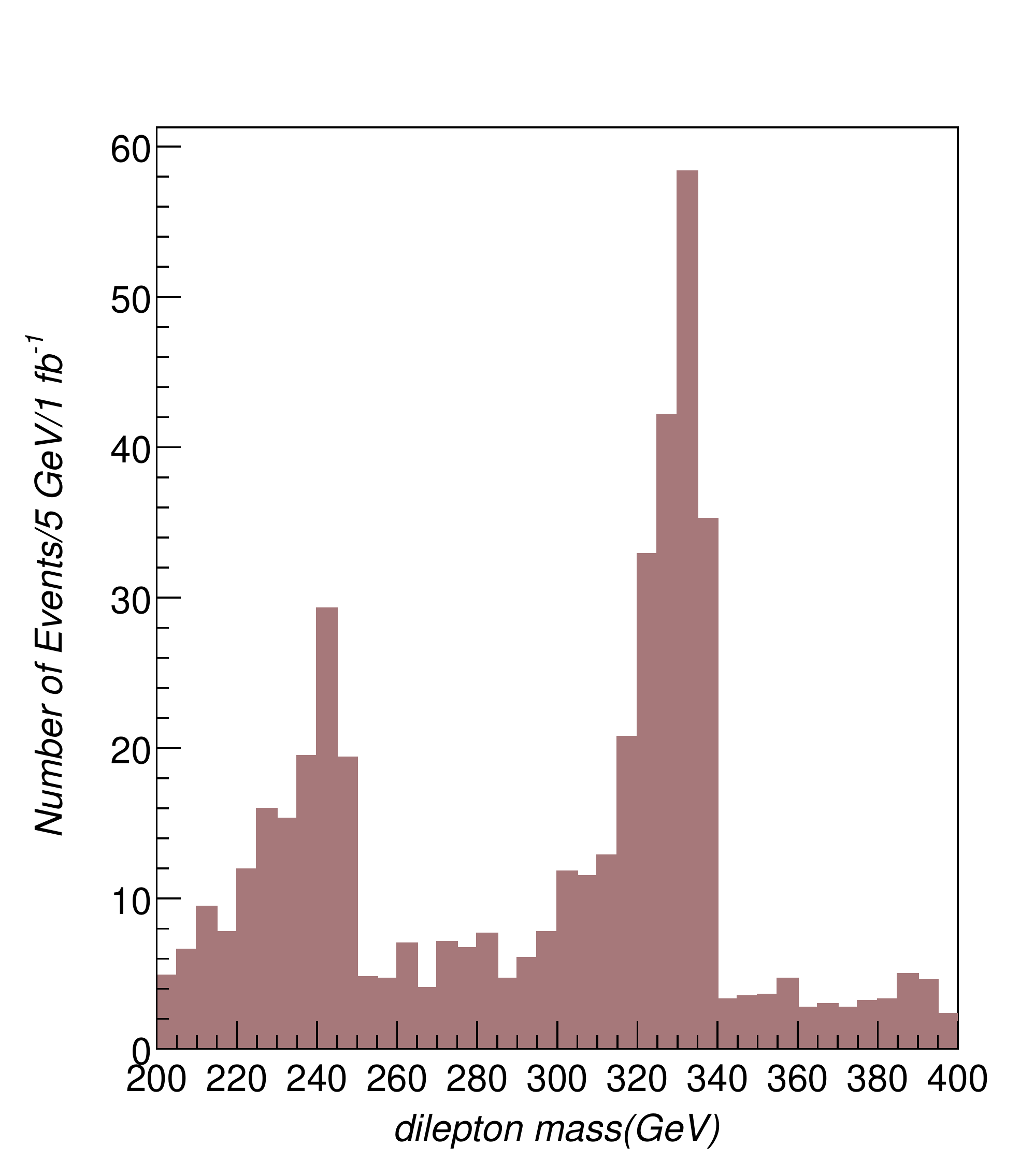}
 \caption{Left: Neutralino reconstruction from slepton-lepton invariant mass for $\sqrt{s}=14 \tev$ and ${\cal L}=1\,\rm{fb}^{-1}$. Right: Dilepton invariant mass distribution.}
\label{2lepneu} 
\end{center}
\end{figure}	

\begin{figure}[t]
\begin{center}
 \includegraphics[scale=0.35]{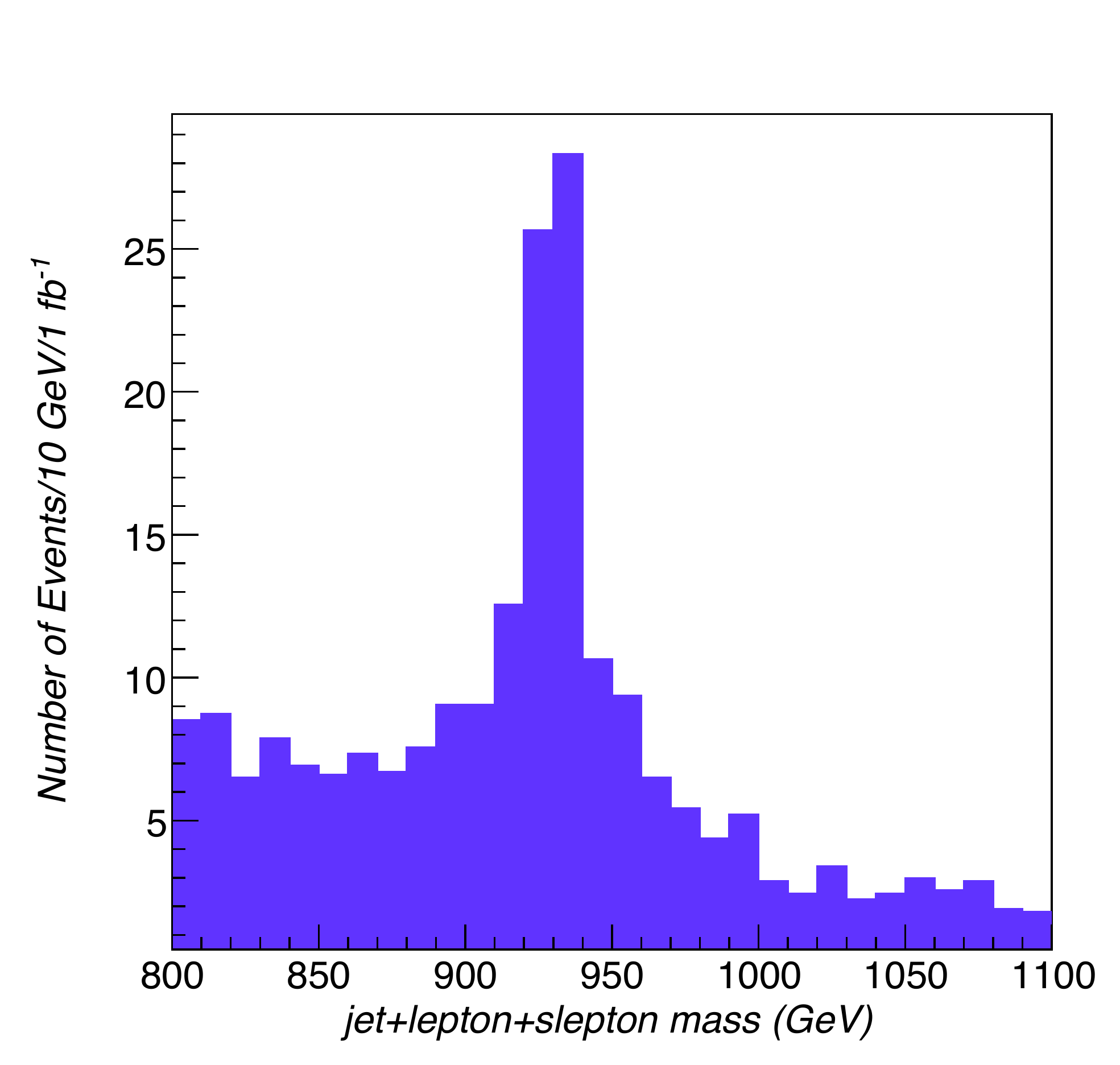}
  \includegraphics[scale=0.4,angle=90]{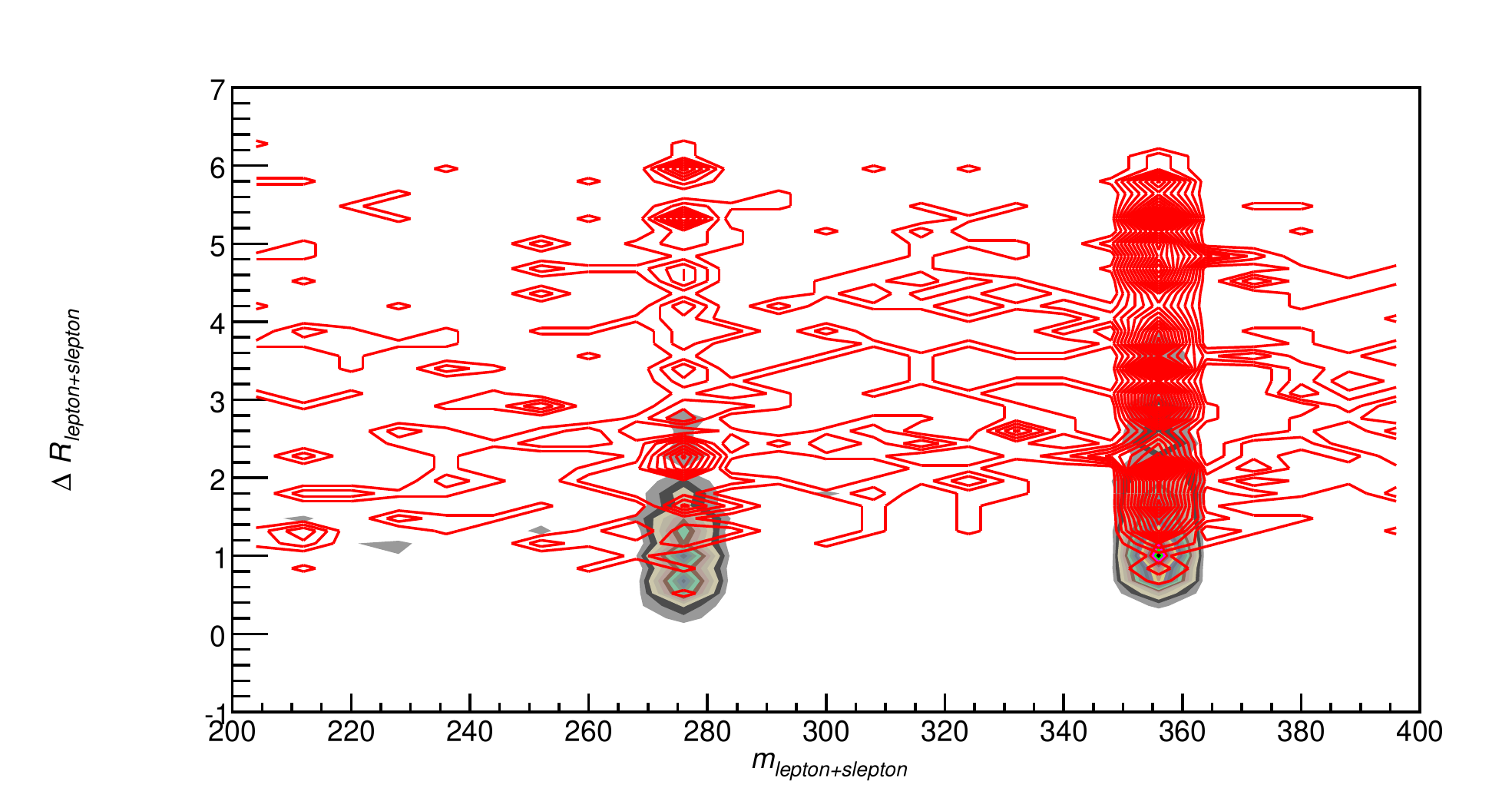}
  \caption{Left: Squark reconstruction for $\sqrt{s}=14 \tev$ and ${\cal L}=1\,\rm{fb}^{-1}$. Right: $\Delta$R distribution for the chosen pair (in contour lines) and the rejected pair (in red).}
\label{2lepsq} 
\end{center}
\end{figure}

\begin{table}[htbp]
  \centering
  \begin{tabular}[c]{ | c | c | c |  c | c |}
  \hline
  &  mass (GeV)   & width (GeV) & Events & Channel \\
  \hline
$\tilde{\chi}_1^0$& 272  & 2    &96 &4\\
$\tilde{\chi}_3^0$& 356  & 2   &132 &4\\
$\tilde{q}$             & 930  & 11 & 83&4 \\
   \hline
$\tilde{\ell}_L$ & $252$  & $3 $ & $68$ &5\\
   \hline  
$\tilde{\chi}_4^0$& $680$ &$10$&$7$ &6\\
   \hline
  \end{tabular}
  \caption{Gaussian fits for  $\sqrt{s}=14 \tev$ and ${\cal{L}}=1\,\rm{fb}^{-1}$. The last column refers to the number of leptons in the channel. For $\sqrt{s}=10 \tev$ and ${\cal{L}}=0.2\,\rm{fb}^{-1}$, each channel event rate is reduced by a factor of 13. }
\label{fit4}
\end{table}

The slepton-lepton invariant mass distribution fully reconstructs the masses of three neutralinos $m_{\tilde{\chi}^0_1}$ and $m_{\tilde{\chi}^0_3}$. We have assumed that enough many slow sleptons are observed to establish the slepton mass. The left plot in Fig.~\ref{2lepneu} exhibits two clear peaks in $m_{\ell \tilde{\ell}}$ corresponding to $\tilde{\chi}_1^0$ and $\tilde{\chi}_3^0$.  The slepton+lepton+jet reconstruction determines $m_{\tilde{q}}$, see Fig.~\ref{2lepsq}. In Table~\ref{fit4} we present the Gaussian fits of the neutralinos and squark masses. On the other hand, for the misidentified sleptons, their energy is taken to be $|\vec{p}|$ where $\vec{p}$ is the three-momentum, which is smaller than their true energy $E=\sqrt{\vec{p}^2+m^2}$. Therefore instead of peaks, the dilepton invariant mass distribution shows edge-like structure with end points at about $\sqrt{m_{\tilde{\chi}^0}^2-m_{\tilde{\ell}_R}^2}$.  Both the shift of the maximum and the smearing of the distribution, apparent on the right plot in Fig.~\ref{2lepneu},
are the result of ``missing mass." The slepton mass is not included in the calculations of energy whenever sleptons are misidentified.

Finally, the difference in $\Delta$R distribution between the chosen dilepton and the wrong dilepton combination is apparent in Fig.~\ref{2lepsq}---{\it correct} pairings have lower $\Delta$R$_{\ell \ell}$. 
 
%%%%%%%%%%
%%%%%%%%%%%%%%%%%%%%%%%%%%%%%%%%%%%%%%%%%%%%%

\subsection{Five-lepton channels}

\begin{figure}[h]
\begin{center}
 \includegraphics[scale=0.8]{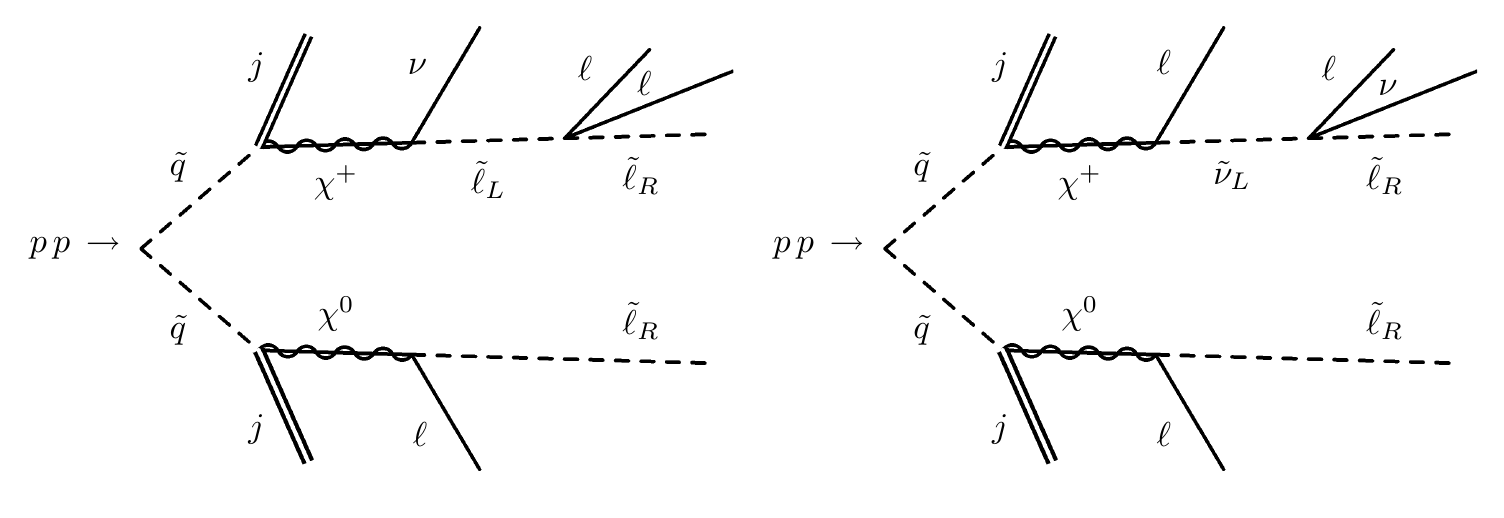}
  \hspace{0.6cm}
    \includegraphics[scale=0.25,angle=90]{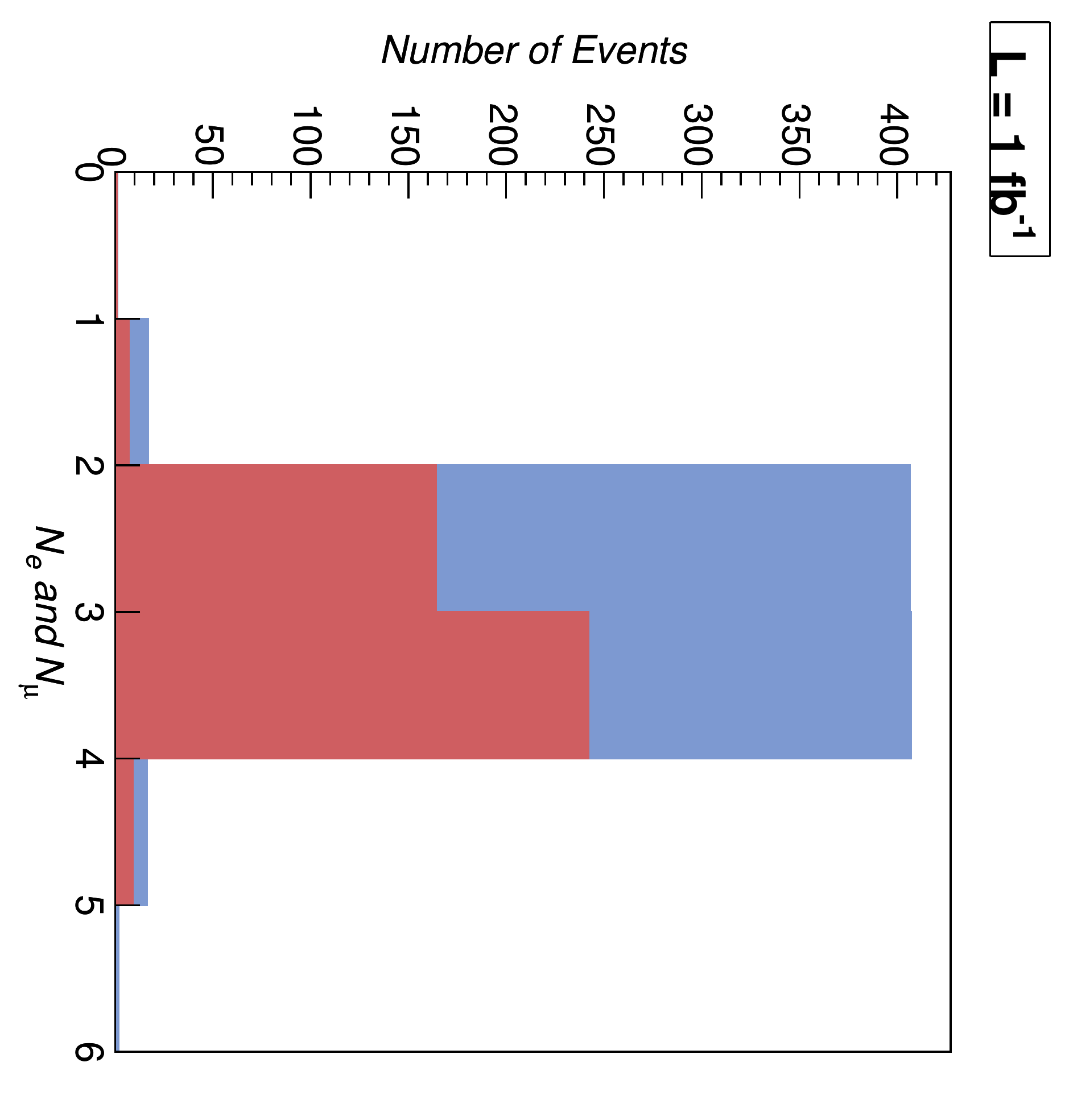}
\caption{Upper: Five-lepton channels. Lower: Electron (lower-red) and muon (upper-blue) composition of the five leptons.}
\label{3lep-feyn} 
\end{center}
\end{figure}

The five-lepton channels, like the ones depicted in Fig.~\ref{3lep-feyn}, have a long decay chain involving $\tilde{\chi}^\pm$ decaying to $\tilde{\ell}_L$ or $\tilde{\nu}$.  Fig.~\ref{3lep-feyn} also shows the muon and electron composition of the five leptons. We use this channel mostly for $\tilde{\ell}_L$ mass reconstruction as well as for gaining a less accurate estimate of the $\tilde{\nu}_L$ and  $\tilde{\chi}^\pm$ masses. 

Events in the five-lepton channel are selected with cuts similar to those used in four-lepton analysis and described in Eq.~(\ref{mincuts}), except
now $n_\ell=5$.  The total cross section of this channel is 137 (426) fb at $\sqrt{s}=10\,(14) \tev$ and gives  27 (426) events at 0.2 (1) fb$^{-1}$.   

\begin{figure}[htbp]
\begin{center}
 \includegraphics[width=6in,height=4in]{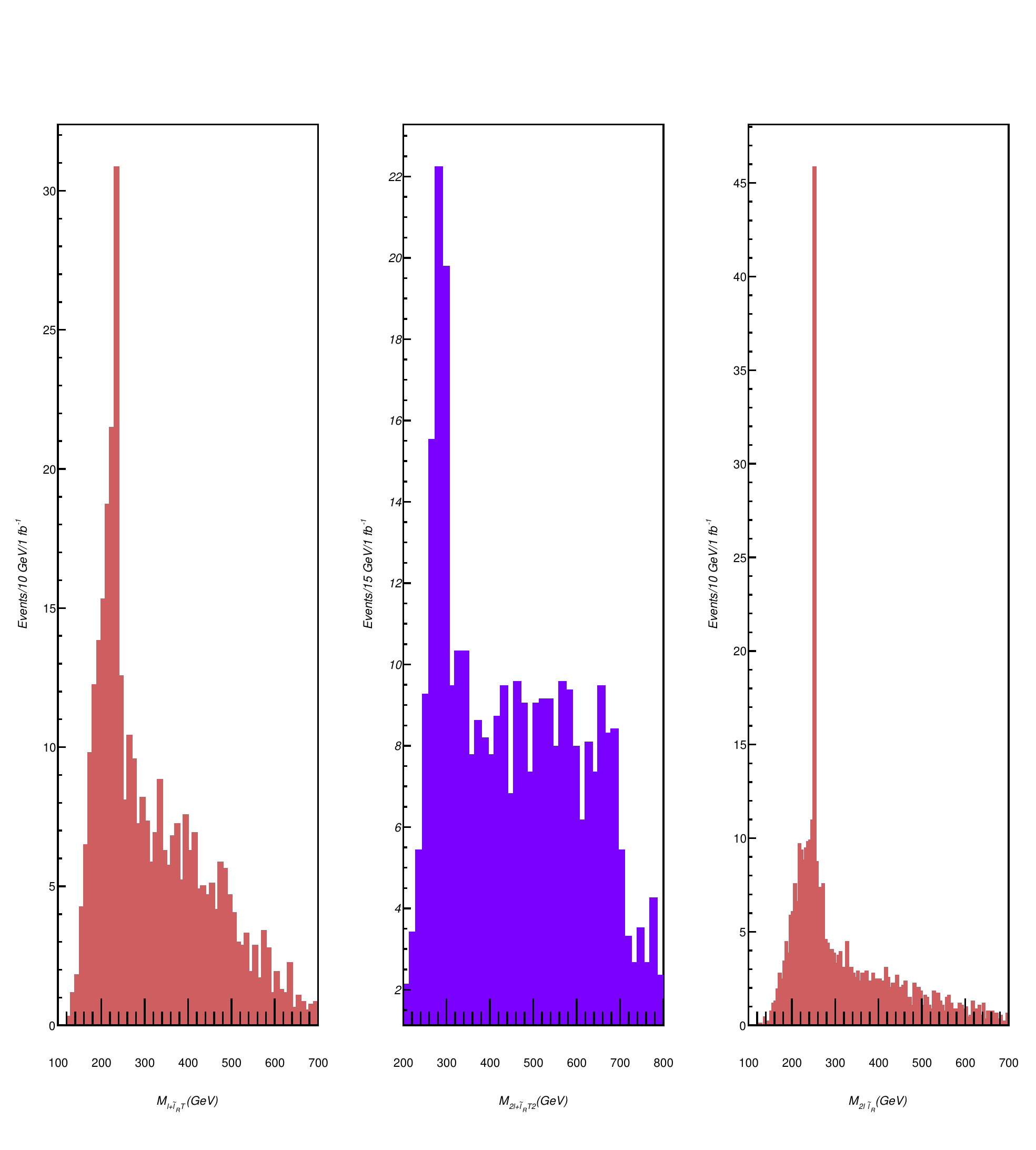}
 \caption{Chargino-sneutrino transverse mass, slepton invariant mass reconstruction for $\sqrt{s}=14 \tev$ and ${\cal L}=1\, \rm{fb}^{-1}$.}
\label{5leprec} 
\end{center}
\end{figure}

For reconstruction, we selected slepton-lepton pairs with smaller $\Delta$R$_{\ell\tilde{\ell}}$ and then formed slepton+2 lepton clusters by adding to the slepton-lepton pair another nearby lepton chosen through $\Delta$R selection.  The decay $\tilde{\ell}_L \to 2 \ell+ \tilde{\ell}_R$ can be fully reconstructed by the slepton+2 lepton invariant mass. The distribution in Fig.~\ref{5leprec} shows a resonance peak at the $\tilde{\ell}_L$ mass, fitted to be $252 \gev$
(see Table~\ref{fit4}). 

The $\tilde{\nu}$ and $\tilde{\chi}^\pm$ cannot be fully reconstructed as their decays involve missing energy.  Instead, we define the transverse mass variables  of the system:
\beq
M_{i \, T}^2&=&(E_{i\, T}+\etmiss)^2-(\vec{p}_{i\, T}+\ptmiss)^2,
 \eeq
where $i$ is either slepton-lepton or slepton+two leptons. We cannot estimate $\etmiss$ mis-calibration in the early running, and the information obtained from transverse 
mass, as in Fig.~\ref{5leprec}, very much depends on how accurate the achieved calibration will be.

\begin{figure}[htbp]
\centering
\includegraphics[scale=0.25]{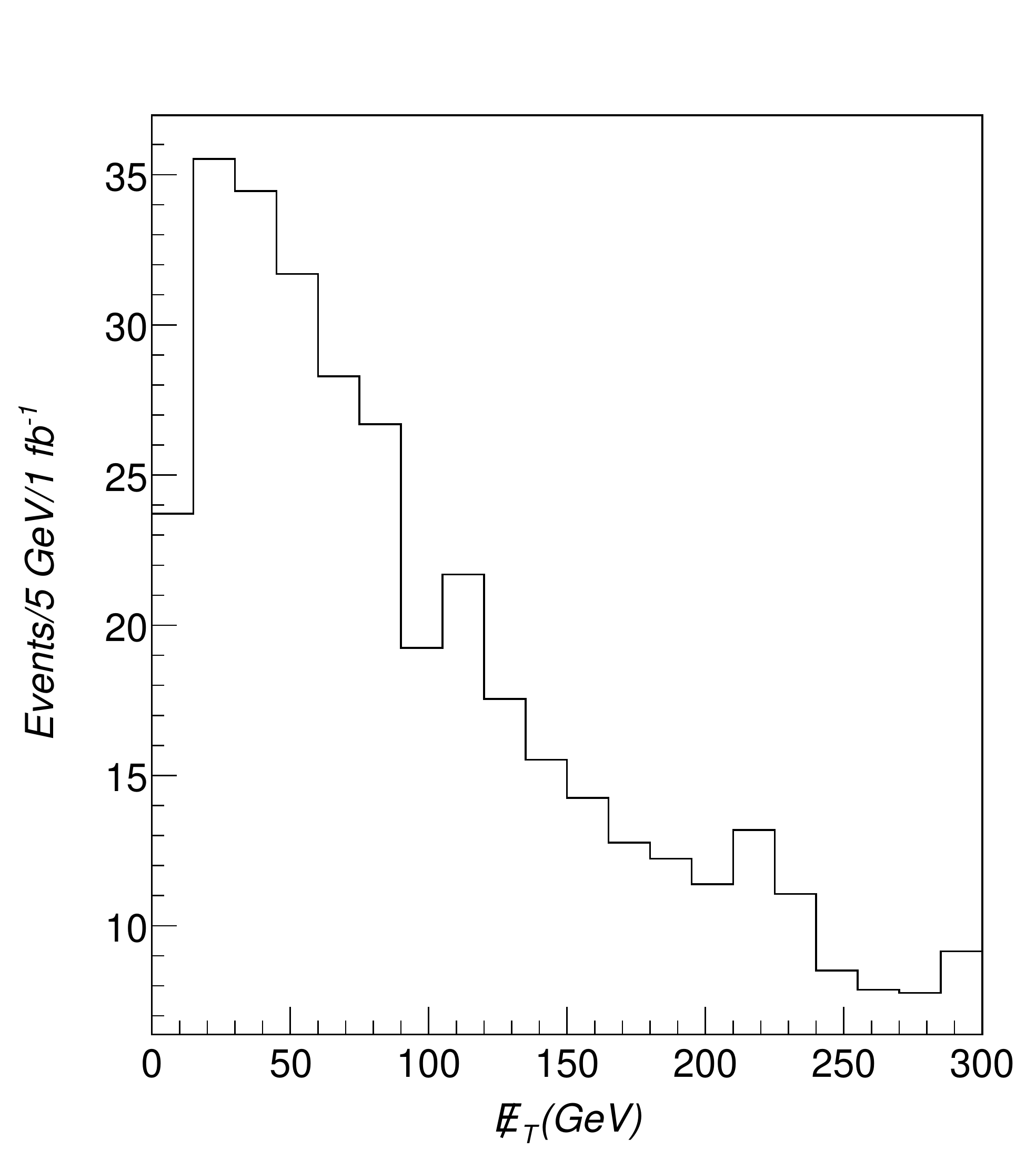}
\includegraphics[scale=0.25]{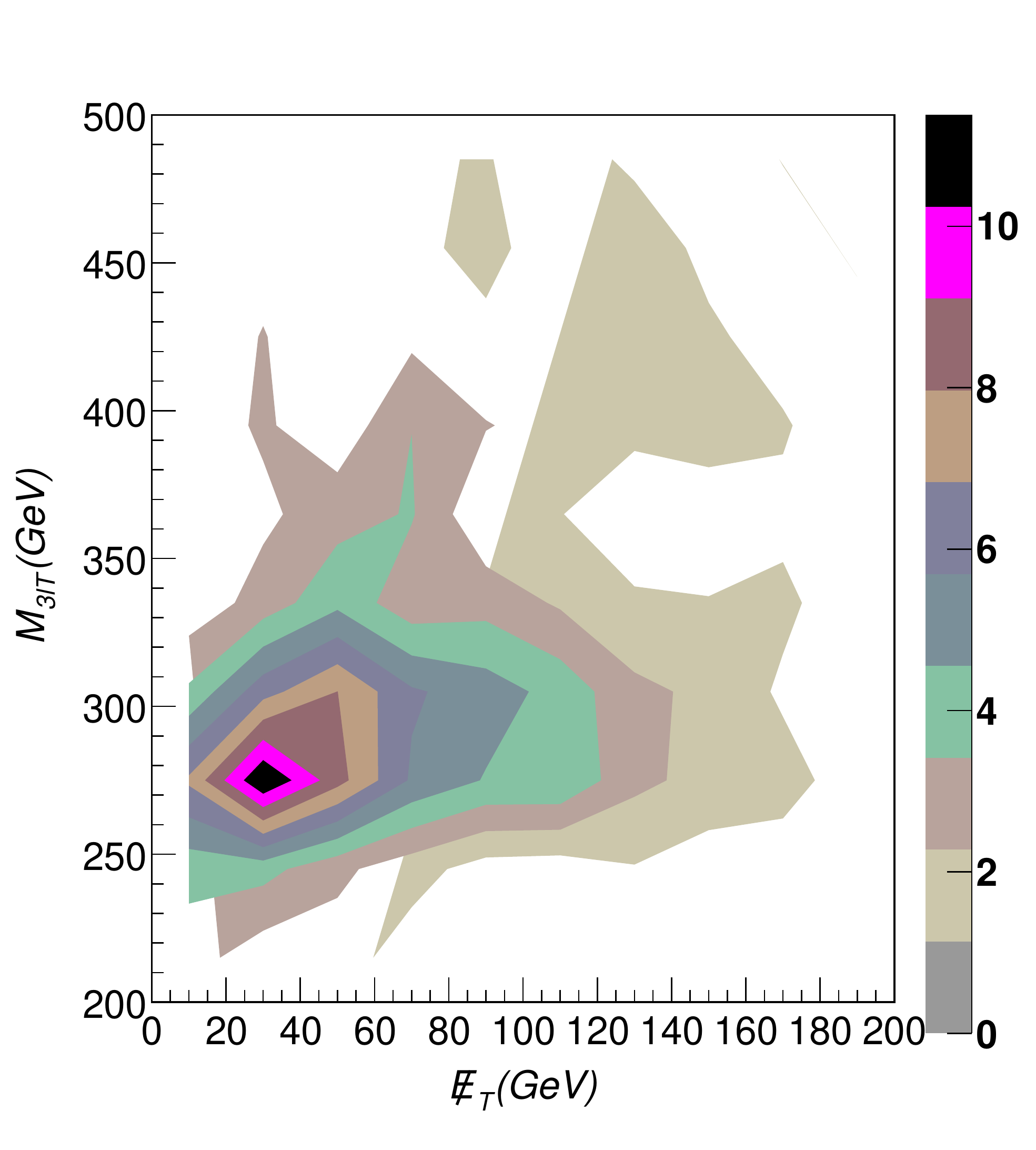}
\caption{ $\etmiss$ distribution and $(\etmiss, M_T)$ distribution.\label{missinget}}
\end{figure}

The missing energy measured by the calorimeter, Fig. \ref{missinget}, is small and is mostly below $100 \gev$. Conservatively, we impose no missing energy cut since selection on the number of leptons is sufficient to reduce backgrounds. Fig.~\ref{missinget} also shows the contours in the $(\etmiss, M_{\tilde{\ell}+2\ell T})$ plane.  The contours are the most dense around $\etmiss \sim 30 \gev$, $M_{\tilde{\ell}+2\ell T}\sim 300 \gev$ corresponding to the $\tilde{\chi}_1^\pm$.    Because the missing energy comes solely from the single neutrino produced by either $\tilde{\nu}$ or $\tilde{\chi}^\pm$ decay, one expects to see Jacobian edges at $M_{\tilde{\ell}+\ell T}\sim M_{\tilde{\nu}}$ and at $M_{\tilde{\ell}+2\ell T}\sim M_{\tilde{\chi}^\pm}$ .  The plots in  Fig.~\ref{5leprec} depict an edge at the sneutrino mass $236 \gev$ in the slepton-lepton transverse mass distribution and two edges at  the chargino masses $294 \gev$ and $677 \gev$ in the slepton+2 lepton  transverse mass distribution.

%%%%%%%%%%%%%%%%%%%%%%%%%%%%%%%%%%%%%%%%%
 %%%%%%%%%%%%%%%%%%%%%%%%%%%%%%%%%%%%%%%%%

\subsection{Six-lepton channels }

Channels with six leptons and no missing energy arise mostly from the combination of two types of cascades
\beq
\textrm{ Short: }\tilde{ \chi}^0 &\to&  \tilde{\ell}_R\,  \ell\, , \\
\textrm{ Long: }\tilde{ \chi}^0 &\to&  \tilde{\ell}_L\,  \ell \to \tilde{\ell'}_R\,   \ell' \, 2 \ell   \, .
\label{chi2nu}
\eeq
In terms of the branching ratios, the dominant cascades are combinations of the long (L) and short (S) cascades.  The (S,L) cascade combination originates from the following gauginos
\begin{eqnarray}
(\tilde{\chi}^0_{1,3} , \tilde{\chi}^0_3), \,  (\tilde{\chi}^0_{1,3}, \tilde{\chi}^0_4), \, ( \tilde{\chi}^0_2, \tilde{\chi}^0_3 ), \, (\tilde{\chi}^0_2, \tilde{\chi}^0_4) ,
\label{SL}
\end{eqnarray}
which are ordered by the total production rate from the highest to lowest. The first neutralino in any bracket yields a short cascade, while the second neutralino a long one.

\begin{figure}[h]
\begin{center}
 \includegraphics[scale=0.6]{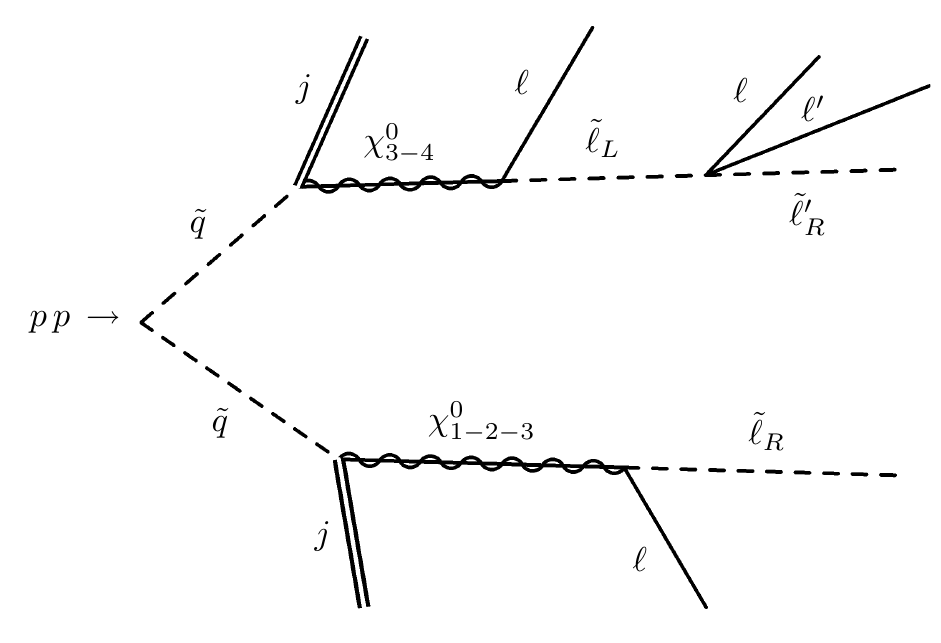}
 \hspace{1.5cm}
   \includegraphics[scale=0.3,angle=90]{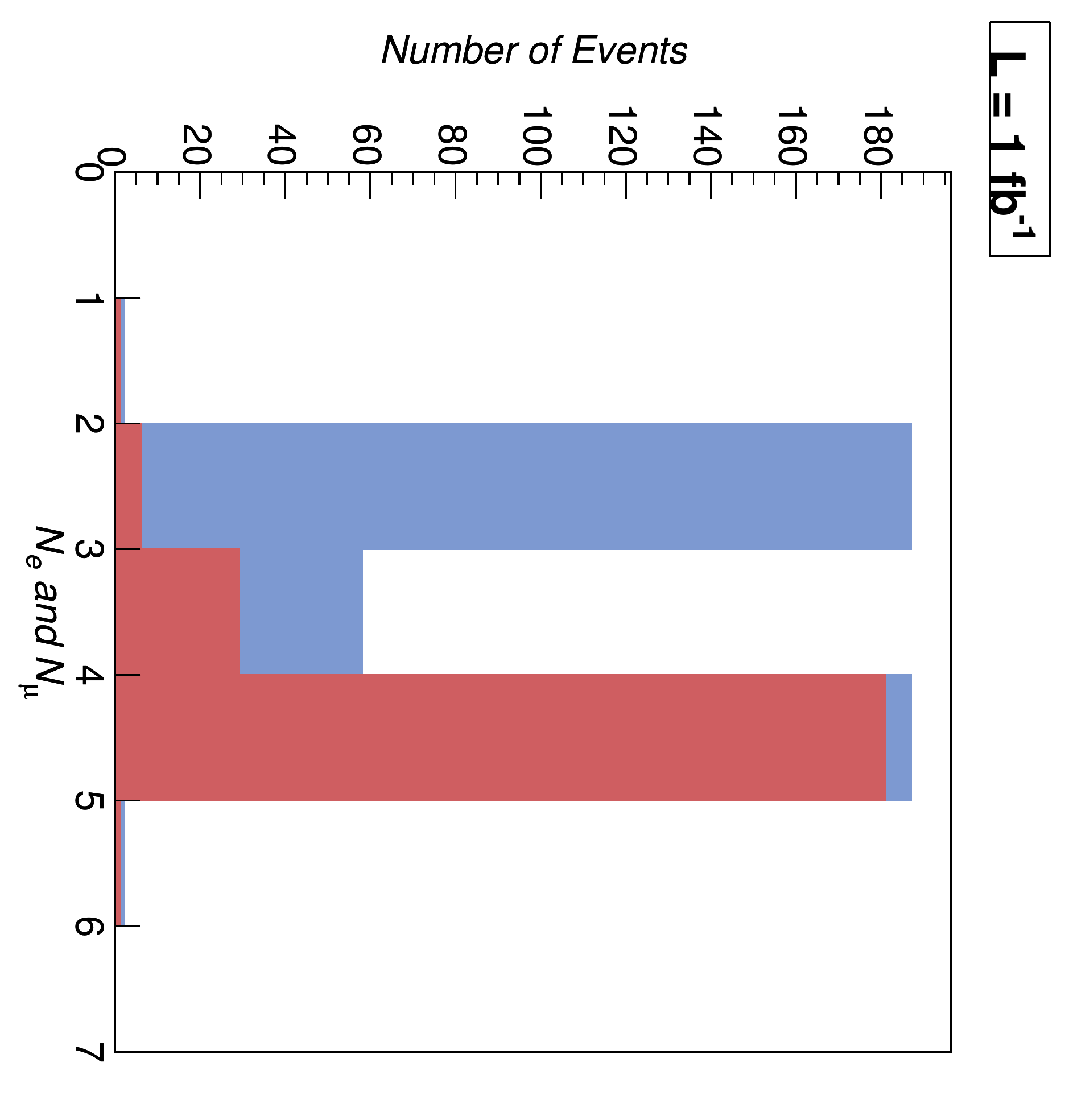}
\caption{Left: Six-lepton channel. Right: Electron (lower-red) and muon (upper-blue) composition of the six leptons.}
\label{6lep-feyn} 
\end{center}
\end{figure}	

There are also some (L,L) combinations producing six leptons and no missing energy, where a tau is identified as a jet. Among those, only one combination $(\tilde{\chi}_4,\tilde{\chi}_4)$ is produced at a comparable rate to that of $(\tilde{\chi}^0_2, \tilde{\chi}^0_3)$, which itself is very rare.
Therefore, our signal mostly comes from a combination of the (S,L) cascades. 

\begin{figure}[t]
\begin{center}
 \includegraphics[width=6in,height=4in]{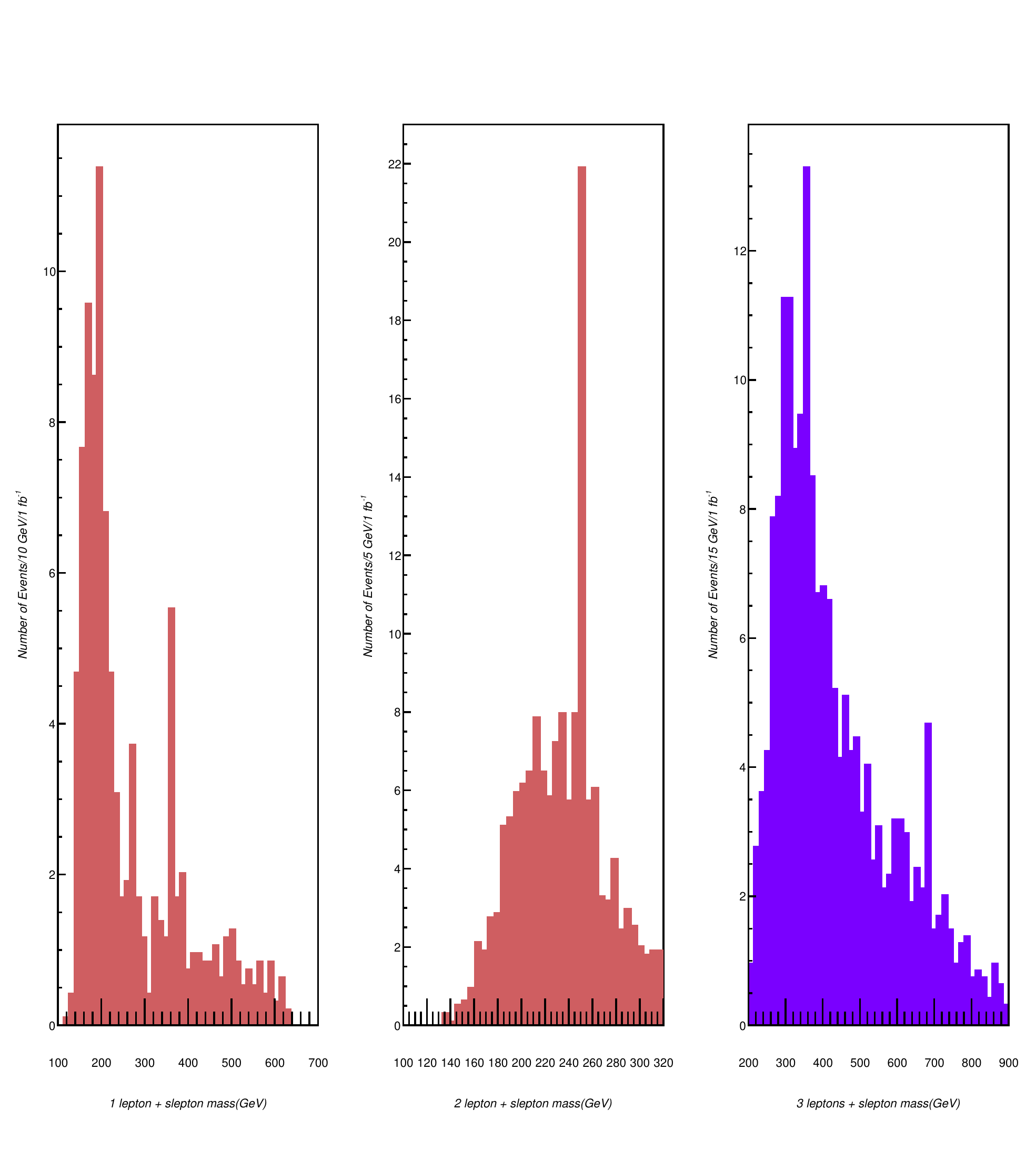}
\caption{Six-lepton channel invariant mass distributions for  $\sqrt{s}=14 \tev$ and ${\cal L}=1\, \rm{fb}^{-1}$ .}
\label{6lep}
\end{center}
\end{figure}

Events in the six-lepton channel are selected with cuts similar to the ones described in Eq.~(\ref{mincuts}), except now $n_{\ell}=6$. The muon and electron composition of the six leptons is shown in Fig.~\ref{6lep-feyn}. The total cross section of this channel is 70 (225) fb at $\sqrt{s}=10\ (14) \tev$ and gives  14 (225) events at 0.2 (1) fb$^{-1}$.  The slepton-lepton invariant mass would reconstruct the neutralino of the short chain with a contamination from partially reconstructing the $\tilde{\ell}_L$ of the long chain. See Fig.~\ref{6lep}. Those neutralinos were already reconstructed in the four-lepton case with better statistics in that channel. 
 
The slepton+2 lepton invariant mass would reconstruct the long chain three-body decay originating from $\tilde{\ell}_L$. The peak corresponding to the slepton mass is clearly visible in Fig.~\ref{6lep}.
 
Using again the $\Delta$R selection, one can pair the combination of the slepton+2 lepton and nearby lepton from the long cascade. The long cascades reconstruct  mostly the $\tilde{\chi}^0_3$ and $\tilde{\chi}^0_4$ since these neutralinos are the most likely ones to originate the long cascades, as described in Eq~(\ref{SL}). Fig.~\ref{6lep} shows clearly the peaks for the $\tilde{\chi}^0_{3,4}$ neutralinos. In Table~\ref{fit4} we quote the Gaussian fit for the $\tilde{\chi}^0_4$ mass. 

Finally, we can pair the neutralino reconstruction with the nearby (in $\Delta$R) leading jet. One can use either the long or the short combination. The squark reconstruction is similar to the four-lepton case but with lower statistics, hence we do not present it here.
  
%%%%%%%%%%%%%%%%%%%%%%%%%%%%%%%%%%%%%%%%%
\subsection{Discovery of the Higgs}
\label{sec:Higgs}

As in any MSSM model, the SM-like Higgs boson in lepto-SUSY is too light to decay into two $W$ bosons.  The Higgs then
decays predominantly into $b\bar{b}$ with a branching ratio of about $80\%$. Due to large backgrounds, the Higgs searches at low mass are focused on a rarer but cleaner decay, $h\to \gamma\gamma$, which greatly limits their statistics.

\begin{figure}[t]
\begin{center}
 \includegraphics[scale=0.7]{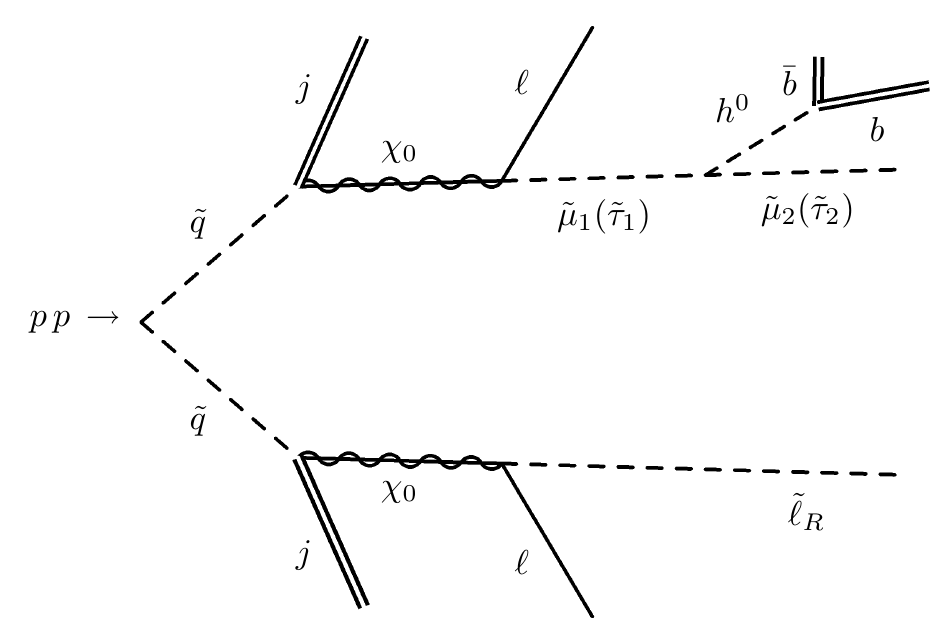}
\caption{Relevant process for Higgs mass reconstruction.}
\label{diagramhiggsbbar}
\end{center}
\end{figure}

In lepto-SUSY, in contrast to common-lore Higgs searches, the Higgs discovery channel is $h\to b\bar{b}$.  Because the Higgs is produced in cascade decays it is free of SM backgrounds
(see discussion in Section \ref{sec4lep} for details on SM backgrounds). It is therefore possible to discover $h$  through the analysis of a clean $b\bar{b}$ invariant mass distribution. The relevant decay chain is illustrated in Fig.~\ref{diagramhiggsbbar}.

Energetic jets, including $b$-jets, are produced in the final state in several ways, not only through the decay of the Higgs boson.
For example,  $\tilde\mu_2$ or $ \tilde{\tau}_2$
at the end of the chain can decay to the $Z$, which in turn can decay to $b\bar b$. For our sample point,
we found that the decays to $h$ are competitive with the decays to $Z$, with  branching ratios:
\begin{eqnarray}
\textrm{BR}(\tilde{\mu}_1\rightarrow h (Z)+\tilde{\mu}_2)
&=&44.1\%\, (35.1\%)\\
\textrm{BR}(\tilde{\tau}_1\rightarrow h (Z)+\tilde{\tau}_2)
&=&53.3\%\, (46.6\%)\,.
\label{BRhZ}
\end{eqnarray}
The Higgs and $Z$ are decayed inside PYTHIA \cite{pythia}.

\begin{figure}[htbp]
\begin{center}
\includegraphics[scale=0.45,angle=90]{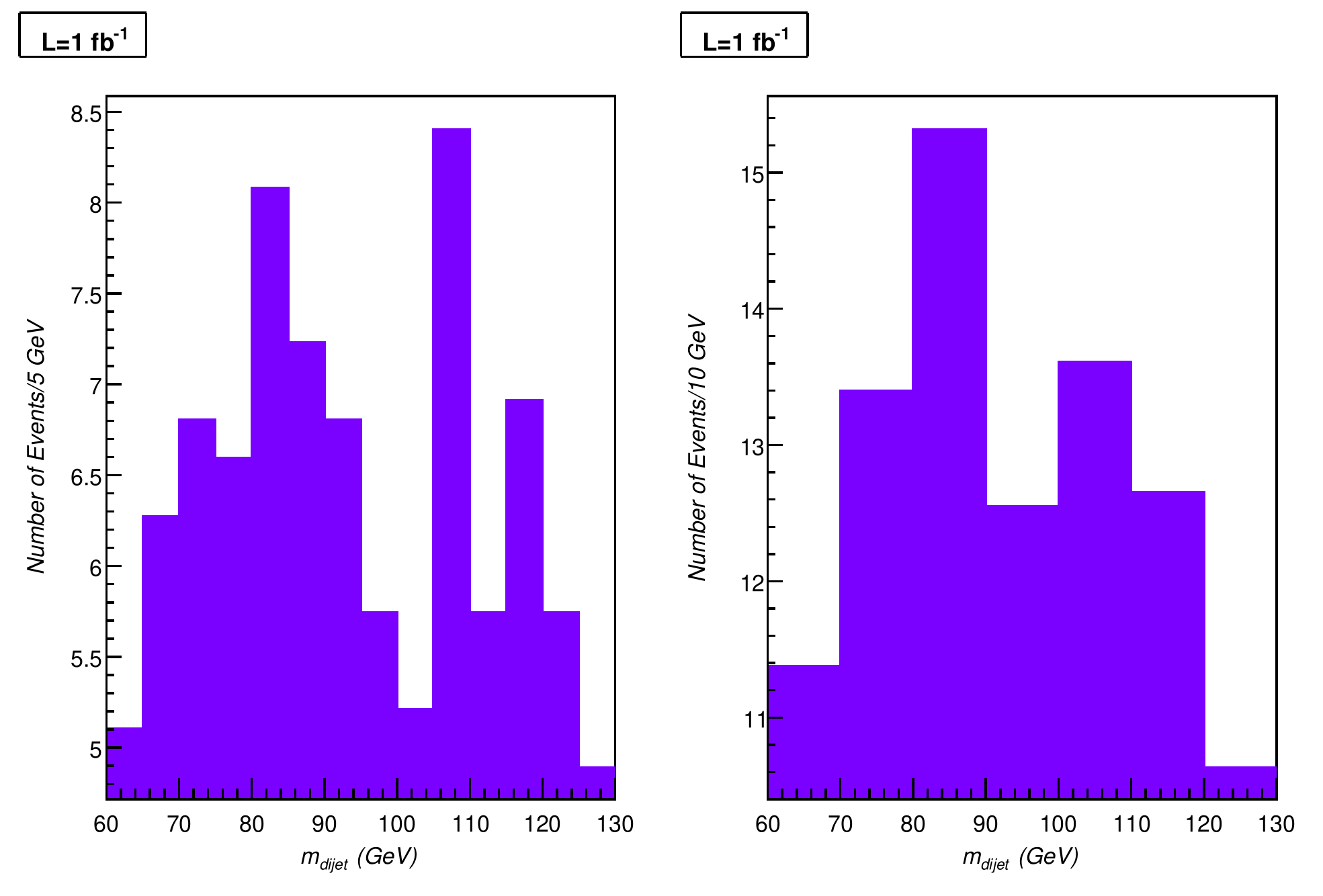}
\caption{Invariant dijet mass distribution with  $\etmiss < 40$~GeV without b-tagging for  $\sqrt{s}=14 \tev$ and ${\cal L}=1 \, \rm{fb}^{-1}$. }
\label{higgsbbar1}
\end{center}
\end{figure}

\begin{figure}[htbp]
\begin{center}
\includegraphics[scale=0.3]{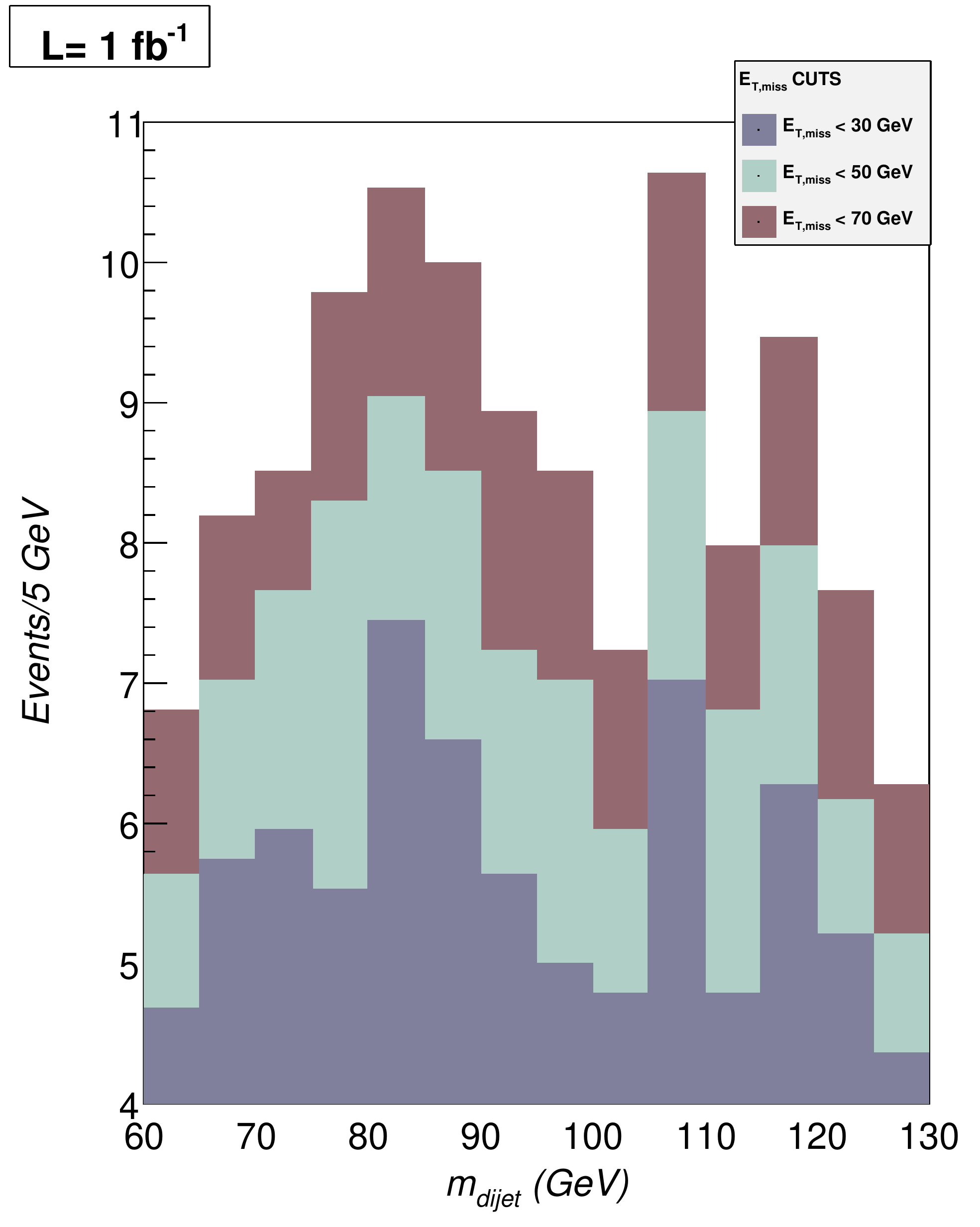}
\caption{Invariant dijet mass distribution for  $\sqrt{s}=14 \tev$ and ${\cal L}=1 \, \rm{fb}^{-1}$ and different $\etmiss$ cuts. }
\label{etcut}
\end{center}
\end{figure}

To reconstruct the Higgs mass, we selected events  characterized by
\beq
n_{l} &\leq &4 \textrm{ with } |\eta|< 2.5, \textrm{ }  p_T> 10 \textrm{ GeV}  \nonumber \\
 &  &\textrm{ and parton level isolation cuts } \Delta R_{\ell\ell}>0.4, \, \Delta R_{\ell j}>0.4, \nonumber \\
n_{j} &\geqslant& 4 \textrm{ with } |\eta|< 2.5, \textrm{ }  p_T> 15  \textrm{ GeV} \nonumber \\
& & \textrm{ and  post-PYTHIA isolation cuts } \Delta R_{jj}>0.4.
\eeq
We ordered the jets according to decreasing $p_T$.
With no b-tagging, we assume the third and fourth jets come from the end of the decay chain and therefore are the b-jets we are interested in. We then construct dijet invariant mass of the third and fourth leading jets to reconstruct the $Z$ and Higgs masses. In Fig.~\ref{higgsbbar1}, the $Z$ and Higgs peaks are clearly visible with bin sizes either of 5 or 10 GeV\@. In Fig.~\ref{etcut}, we show how relaxing the $\etmiss$ cut slowly introduces more combinatorics but still retains proportionately the $Z$ and $h$ signals. We chose a cut $\etmiss< 40$ GeV, which we consider a conservative choice for early running.
The total cross section for those events is  100 (320) fb at 10 (14) TeV, which leads to about 20 events in the 200 pb$^{-1}$ run at 10 TeV.

We can use rough cuts on the invariant mass to estimate the significance of these events. With 1 fb$^{-1}$ of data, there are 50 events for  $m_{\rm{dijet}}$ between 60 and 100 GeV and 37 events  between 100 GeV and 130 GeV. But that counting does not take into account two important facts. First, the $Z$ and $h$ peaks overlap. We used Gaussian fits to estimate that the Higgs peak contains 51 events and the $Z$ contamination under the $h$ Gaussian is 16 events, leading to 35 Higgs events at 1 fb$^{-1}$ (see Table~\ref{hfit} for details). The second important effect is the combinatorial background which lies below the two peaks, and extended in a larger range of dijet masses. The combinatorial background increases by relaxing the $\etmiss$ cut  and could be fitted with a parabola or a gaussian. Our conclusion is that a proper treatment of the combinatorial background would require full simulation.

\begin{table}[h]
 \centering
 \begin{tabular}[c]{ | c | c | c |  c | }
 \hline
 &   mass (GeV)   & width (GeV) & Events \\
 \hline
$Z$& 82   & 23 & 82 \\
$h$ & 113  &15 & 51  \\
\hline
 \end{tabular}
 \caption{Fits of the $Z$ and $h$ peaks with $\etmiss < 40$~GeV, for  $\sqrt{s}=14 \tev$ and ${\cal{L}}=
 1\,\rm{fb}^{-1}$. }
 \label{hfit}
\end{table}

%%%%%%%%%%%%%%%%%%%%%%%%%%%%%%%%%%%%%%%%%
\subsection{Dirac or Majorana gauginos}
\label{dirac}
Lepto-SUSY spectrum can arise in models that contain an approximate $U(1)_R$ symmetry, making Dirac gaugino masses necessary.
Using the clean lepto-SUSY channels, it is straightforward to distinguish the Majorana or Dirac nature of the neutralinos.
For example, a Dirac gluino forbids squark pair production channels, like $pp \to \tilde{q}\tilde{q}$, while $pp\to \tilde{q}\bar{\tilde{q}}$ is allowed \cite{Fox:2002bu,diracphe}. In the cleanest four-lepton final state in lepto-SUSY we have simulated, the selected events have both same-sign and opposite-sign leptons. By simple combinatorics, both configurations have the same probabilities. In the Dirac gaugino case, only the opposite-sign lepton configuration remains. 
The lepton charge distribution gives a straightforward discriminant of the two types of gaugino soft mass.

%%%%%%%%%%%%%%%%%%%%%%%%%%%%%%%%%%%%%%%%%
%%%%%%%%%%%%%%%%%%%%%%%%%%%%%%%%%%%%%%%%%
%%%%%%%%%%%%%%%%%%%%%%%%%%%%%%%%%%%%%%%%%
\section{Models with lepto-SUSY spectrum}
\label{sec:models}

In this section we propose a model-independent parametrization of the lepto-SUSY soft mass terms and  discuss models that lead to lepto-SUSY spectra. Our parameterization is general, but it is most useful for models in which the supersymmetry mediation mechanism inolves gauge interactions. 

We parametrize the soft mass terms assuming that various superpartners acquire soft mass contributions proportional to their SM gauge charges. The scalar masses pick up contributions from every gauge interaction they participate in, where each gauge group is characterized by the corresponding coefficient $K_i$
\beq
  \tilde{m}^2 (R) = \sum_{i=1}^3 C_2(R_i) K_i \, ,
  \label{eq:softmass}
\eeq
where the sum runs over the SM gauge groups and the quadratic Casimir, $C_2$, is $(N^2-1)/2N$
for an $SU(N)$ fundamental representation while it is $3/5$ times the hypercharge squared for $U(1)$. The $R_i$ denote the representations of the field
under the corresponding SM gauge groups. Finally, the coefficients $K_i$'s encode the details of the particular supersymmetry breaking mechanism in play,
\beq
 K_i = \frac{\alpha_i}{\pi}m_i^2n_i^2 \qquad (i=1,2,3),
\eeq
where $m_i$ are the gaugino masses. 

With unspecified dimensionless $n_i$'s, this parametrization is still completely general subject only to the assumption in Eq.~(\ref{eq:softmass}). In specific models, like GMSB~\cite{gmsb} with large number of messengers or L$\tilde{g}$M~\cite{lgm} (see also~\cite{Chacko:2003tf},~\cite{Medina:2006hi}), it turns out that the parameters $n_i$ are ${\cal{O}}(1-10)$ numbers. Our parameterization stresses the fact that scalar soft masses are typically one-loop suppressed compared to the corresponding gaugino masses squared in such models. The squares of the parameters $n_i$ are smaller than the loop suppression and therefore the scalar soft masses are smaller, by a factor of a few, than the gaugino masses. 

One additional parameter needs to be introduced in the Higgs sector. According to Eq.~(\ref{eq:softmass}),  $m_{H_d}^2$ and $m_{H_u}^2$ are both positive. Typically, EWSB is triggered by additional contributions to $m_{H_u}^2$ induced by the large value of the top quark Yukawa coupling. We incorporate such a contribution as a free parameter $n_4$: 
\beq 
\label{eq:Higgsdelta}
    \delta\equiv -m_{H_d}^2+m_{H_u}^2&=&-\frac{\alpha_3\lambda_t^2}{4\pi^3}m_3^2n_4^2 \, .
\eeq 

We also assume that the gaugino masses obey the unified relations, that is $m_{i} \propto \alpha_i$. In summary, our parameter space is defined by an overall scale, 
given by the gluino mass, four dimensionless parameters which parametrize the details of SUSY breaking mechanism, $\tan \beta$, and $\rm{sign} \,\mu$:
\beq 
m_3,\,n_i\,(i=1,2,3,4),\,\tan \beta, \, \rm{sign}\,
\mu.
\eeq
We neglect the trilinear A-terms as they tend to be small in the GMSB and L$\tilde{g}$M. As we discussed earlier, the A-terms would have to be unnaturally large to make a sizable impact on the mass spectrum.

In Section~\ref{sec:setup} we discussed the lepto-SUSY spectrum and constraints on the spectrum. In terms of our parameterization, the mass ranges in Eq.~(\ref{eq:massrange}), correspond to
\begin{center}
\begin{tabular}{cc}Parameter & Range \\ $n_1$ & [2, 5] \\ $n_2$ & [0.5, 6] \\ $n_3$ & $>1.8$ \\ $n_4$ & $>1.75$\end{tabular}
 \end{center}
Therefore, lepto-SUSY spectra are obtained for ${\cal O}(1)$ parameters. For example, the sample spectrum presented in Table~\protect\ref{table:sample} is obtained from the input parameters
in Table~\ref{table:input}.
\begin{table}[h!tb]
  \centering
  \begin{tabular}[c]{ c | c |  c | c |}
  \hline
   \multicolumn{2}{|c|} {$m_3 $ }   &  \multicolumn{2}{|c|} {$2000 \gev$ }  \\
    \multicolumn{2}{|c|} {$n_1 $ }   &  \multicolumn{2}{|c|} {$4.8$ }  \\
 \multicolumn{2}{|c|} {$n_2 $ }   &  \multicolumn{2}{|c|} {$3.9$ }  \\
  \multicolumn{2}{|c|} {$n_3 $ }   &  \multicolumn{2}{|c|} {$2.2$ }  \\
   \multicolumn{2}{|c|} {$n_4 $ }   &  \multicolumn{2}{|c|} {$6.7$ }  \\
   \multicolumn{2}{|c|} {$\tan \beta$} &
         \multicolumn{2}{|c|} {$10$}   \\
          \multicolumn{2}{|c|} {$\rm{sign}\ \mu$} &
           \multicolumn{2}{|c|} {$ +$}  \\
  \hline
  \end{tabular}
  \caption{Input parameters that correspond to the sample spectrum in Table~\protect\ref{table:sample}.   \label{table:input} }
\end{table}

The parameters $n_i$ are ${\cal O}(1)$ numbers in models where there is a suppression of the squark and slepton masses compared to the gaugino masses of order of a loop factor. For illustration, we briefly examine how such ${\cal O}(1)$ values of $n_i$ arise in some models. 

In GMSB, the gaugino masses are generated at one-loop level, while the scalar masses squared at two loops. Both the gaugino and scalar masses squared are proportional to the number of messengers, $N_{m}$. Parametrically
\beq
  n_i\propto \frac{1}{\sqrt{N_m}} \sqrt\frac{\pi}{\alpha_i}\, ,
\eeq
that is one obtains lepto-SUSY spectrum when the number of messengers is of order of the loop suppression factors. The above equation holds for $n_4$ as well since soft Higgs masses are generated at the next loop order compared to the scalar masses. 

In gaugino-mediation~\cite{Kaplan:1999ac}, the scalar masses are generated radiatively through diagrams involving the gauginos. Whether or not the scalar masses are suppressed compared to the gauginos depends on the mediation scale. In high-scale models~\cite{highscale}, the natural one-loop suppression is compensated for by large logarithm of the ratio of the mediation scale to the weak scale. In L$\tilde{g}M$  one naturally obtains $n_i\approx 1$. The $m_{H_d}^2 - m_{H_u}^2$ mass splitting, parameterized by $n_4$, is generated by two-loop diagrams involving the stop and is of order indicated in Eq.~(\ref{eq:Higgsdelta}). 

Another example of models with lepto-SUSY spectra are models with supersoft SUSY breaking~\cite{Fox:2002bu}.  In such models, SUSY breaking is caused by a D-term vev in a new $U(1)$ gauge sector. Due to unbroken R-symmetry the gaugino masses are necessarily Dirac type.  The SUSY-breaking D-term couples to the visible sector through higher dimensional operators. The lowest dimensional operator responsible for gaugino masses is a dimension five operator, while the operator responsible for scalar masses is of dimension ten. The suppression of the scalar masses compared to that of the gauginos is therefore natural for such models. Dirac or Majorana gaugino masses can be easily distinguished using charge asymmetries as explained in Section~\ref{dirac}.

%%%%%%%%%%%%%%%%%%%%%%%%%%%%%%%%%%%%%%%%%
%%%%%%%%%%%%%%%%%%%%%%%%%%%%%%%%%%%%%%%%%
%%%%%%%%%%%%%%%%%%%%%%%%%%%%%%%%%%%%%%%%%

\section{Conclusion and outlook}
\label{concl}
Lepto-SUSY is a well-motivated supersymmetric scenario. The ordering of the spectrum ensures, due to kinematics alone, copious lepton production in decay chains.
Therefore, the signals are very clean and the LHC has a tremendous discovery potential. Another interesting feature is that the Higgs boson is produced in SUSY cascade decays,
which reduce the background so much that the Higgs can be discovered via the $b\bar{b}$ channel.

We studied a sample spectrum of lepto-SUSY and estimated the LHC potential of discovering and reconstructing this type of spectrum.
The discovery channels have cross sections governed by the QCD production of squarks and are characterized by two hard jets, two collider-stable sleptons, and at least two leptons.
If the sleptons are misidentified as muons, which is likely,  the signatures are four or more leptons and two hard jets.
A discovery of the stable slepton is possible in the early running of the LHC. Lepto-SUSY with 1~TeV squarks generates sleptons with velocities $0.6<\beta<0.8\ (0.9) $ at a rate of 210 (400) fb at 10 TeV, resulting in 40 (80) stable sleptons with 200 pb$^{-1}$ of data.

The Higgs  discovery channel is  through the $h\to b \bar{b}$ decay in association with four leptons and two hard jets. We estimate the  prospects for Higgs discovery are good with less than 1 fb$^{-1}$ of data at 14 TeV, again assuming 1~TeV squarks.

Using the four-lepton channel one can determine the squark masses and all neutralinos, except the second Higgsino and Wino, in the 10 TeV run with 200 pb$^{-1}$ of data. In the five-lepton channel---which is the only $\etmiss$ channel---one could further determine the masses of the sneutrino, heavy slepton, and charginos using the transverse mass variables. The six-lepton channel has a smaller branching ratio, but it will allow one to determine the Wino mass with just 1 fb$^{-1}$ of data during the 14 TeV run.

There are several directions worth pursuing in the context of lepto-SUSY. The light Higgs is produced in a  clean environment, so one may attempt to use it for an extraction of the bottom Yukawa coupling. We only studied the light Higgs, but there may be new ways to discover the remaining Higgs bosons in lepto-SUSY. We did not investigate how changing the spectrum, by altering various mass ratios, affects the signatures of interest.  For instance, increasing the squark-gluino mass ratio can enhance the squark-gluino associated production, which in turn, may offer new signatures and, of course, prospects for the gluino mass measurement.  We have not fully exploited the flavor information in our analysis.
Since the collider-stable sleptons will often be misidentified as muons one could take advantage of this fact to refine the analysis. Last but not least, for the channels with $\etmiss$ one could use more sophisticated kinematic variables to improve the sensitivity.

\label{conclusions}

%%%%%%%%&&&&&&&&&
%%%%%%%%%%%%%%%%%%%%%%%%%%%%%%%%%%%%%%%%%
%%%%%%%%%%%%%%%%%%%%%%%%%%%%%%%%%%%%%%%%%
%%%%%%%%%%%%%%%%%%%%%%%%%%%%%%%%%%%%%%%%%

\acknowledgments

We thank Martin Schmaltz for collaboration in the early stages of this work and useful suggestions. We would like to thank Ketevi Assamagan, Tulika Bose, Adam Martin for their comments and suggestions. We also thank Steve Martin for pointing out an error in an earlier version of this analysis. The work of ADS is supported in part by the INFN ``Bruno Rossi'' Fellowship and in part  by the US Department of Energy (DoE) under contract No.~DE-FG02-05ER41360. 
JF  and WS are supported in part by the DoE under grant No. DE-FG-02-92ER40704, while VS under grant No. DE-FG02-91ER40676. VS and WS would like to thank the Aspen Center of Physics, where part of this work was done.

%%%%%%%%%%%%%%%%%%%%%%%%%%%%%%%%%%%%%%%%%
%%%%%%%%%%%     BEGIN BIBLIOGRAPHY         %%%%%%%%%%%%%
%%%%%%%%%%%%%%%%%%%%%%%%%%%%%%%%%%%%%%%%%


\begin{thebibliography}{99}
\footnotesize{

%\cite{DeSimone:2008gm}
\bibitem{lgm}
  A.~De Simone, J.~Fan, M.~Schmaltz and W.~Skiba,
  %``Low-scale gaugino mediation, lots of leptons at the LHC,''
  Phys.\ Rev.\  D {\bf 78}, 095010 (2008)
  [arXiv:0808.2052 [hep-ph]].
  %%CITATION = PHRVA,D78,095010;%%




%\cite{Meade:2008wd}
 \bibitem{gmsb}
G.~F.~Giudice and R.~Rattazzi,
 ``Theories with gauge-mediated supersymmetry breaking,''
 Phys.\ Rept.\  {\bf 322}, 419 (1999)
 [arXiv:hep-ph/9801271].
  %%CITATION = PRPLC,322,419;%%


\bibitem{Fox:2002bu}
 P.~J.~Fox, A.~E.~Nelson and N.~Weiner,
 ``Dirac gaugino masses and supersoft supersymmetry breaking,''
 JHEP {\bf 0208}, 035 (2002)
 [arXiv:hep-ph/0206096];
 %%CITATION = JHEPA,0208,035;%%
%\bibitem{Nelson:2002ca}
 A.~E.~Nelson, N.~Rius, V.~Sanz and M.~Unsal,
 ``The minimal supersymmetric model without a mu term,''
 JHEP {\bf 0208}, 039 (2002)
 [arXiv:hep-ph/0206102];
   %%CITATION = JHEPA,0208,039;%%
 Z.~Chacko, P.~J.~Fox and H.~Murayama,
  ``Localized supersoft supersymmetry breaking,''
  Nucl.\ Phys.\  B {\bf 706}, 53 (2005)
  [arXiv:hep-ph/0406142].
   %%CITATION = NUPHA,B706,53;%%
 

\bibitem{cernsearch}

 J.~Abdallah {\it et al.}  [DELPHI Collaboration],
  ``Search for supersymmetric particles in light gravitino scenarios and sleptons NLSP,''
  Eur.\ Phys.\ J.\  C {\bf 27}, 153 (2003)
  [arXiv:hep-ex/0303025].
    %%CITATION = EPHJA,C27,153;%%


\bibitem{susyhit}
 A.~Djouadi, M.~M.~Muhlleitner and M.~Spira,
 ``Decays of Supersymmetric Particles: the program SUSY-HIT
 (SUspect-SdecaY-Hdecay-InTerface),''
 Acta Phys.\ Polon.\  B {\bf 38}, 635 (2007)
 [arXiv:hep-ph/0609292];
 %%CITATION = APPOA,B38,635;%%
A.~Djouadi, J.~L.~Kneur and G.~Moultaka,
  ``SuSpect: A Fortran code for the supersymmetric and Higgs particle spectrum in the MSSM,''
  Comput.\ Phys.\ Commun.\  {\bf 176}, 426 (2007)
  [arXiv:hep-ph/0211331].
 %%CITATION = CPHCB,176,426;%%


\bibitem{primer}
S.~P.~Martin,
``A Supersymmetry Primer,''
 arXiv:hep-ph/9709356;
 %%CITATION = HEP-PH/9709356;%%
 T.~Plehn,
  ``LHC Phenomenology for Physics Hunters,''
  arXiv:0810.2281 [hep-ph];
    %%CITATION = ARXIV:0810.2281;%%
B.~C.~Allanach,
  ``SUSY Predictions and SUSY Tools at the LHC,''
  Eur.\ Phys.\ J.\  C {\bf 59}, 427 (2009)
  [arXiv:0805.2088 [hep-ph]].
 %%CITATION = EPHJA,C59,427;%%
 
   \bibitem{Ambrosanio}
 S.~Ambrosanio, G.~D.~Kribs and S.~P.~Martin,
 ``Three-body decays of selectrons and smuons in low-energy supersymmetry
 breaking models,''
 Nucl.\ Phys.\  B {\bf 516}, 55 (1998)
 [arXiv:hep-ph/9710217].
 %%CITATION = NUPHA,B516,55;%%

 \bibitem{atlas-tdr}
``ATLAS detector and physics performance. Technical design report.  Vol. 2,''
New: ``Expected Performance of the ATLAS Experiment : Detector, Trigger and Physics", arXiv:0901.0512 ; CERN-OPEN-2008-020.
 %%CITATION = ARXIV:0901.0512;%%


\bibitem{thomas}
 S.~Ambrosanio, S.~Heinemeyer, B.~Mele, S.~Petrarca, G.~Polesello, A.~Rimoldi and G.~Weiglein,
  ``Aspects of GMSB phenomenology at TeV colliders,''
  arXiv:hep-ph/0002191;
   %%CITATION = HEP-PH/0002191;%%
S.~Dimopoulos, S.~D.~Thomas and J.~D.~Wells,
  ``Sparticle spectroscopy and electroweak symmetry breaking with gauge-mediated supersymmetry breaking,''
  Nucl.\ Phys.\  B {\bf 488}, 39 (1997)
  [arXiv:hep-ph/9609434];
   %%CITATION = NUPHA,B488,39;%%
  S.~Dimopoulos, M.~Dine, S.~Raby, S.~D.~Thomas and J.~D.~Wells,
 ``Phenomenological implications of low energy supersymmetry breaking,''
  Nucl.\ Phys.\ Proc.\ Suppl.\  {\bf 52A}, 38 (1997)
  [arXiv:hep-ph/9607450];
    %%CITATION = NUPHZ,52A,38;%%
 S.~Dimopoulos, M.~Dine, S.~Raby and S.~D.~Thomas,
  ``Experimental Signatures of Low Energy Gauge Mediated Supersymmetry Breaking,''
  Phys.\ Rev.\ Lett.\  {\bf 76}, 3494 (1996)
  [arXiv:hep-ph/9601367];
    %%CITATION = PRLTA,76,3494;%%


\bibitem{ellis}
 J.~R.~Ellis, A.~R.~Raklev and O.~K.~Oye,
  ``Gravitino dark matter scenarios with massive metastable charged  sparticles at the LHC,''
  JHEP {\bf 0610}, 061 (2006)
  [arXiv:hep-ph/0607261].
 %%CITATION = JHEPA,0610,061;%%




\bibitem{HaberKane}
 H.~E.~Haber and G.~L.~Kane,
 ``The Search For Supersymmetry: Probing Physics Beyond The Standard Model,''
 Phys.\ Rept.\  {\bf 117}, 75 (1985).
%%CITATION = PRPLC,117,75;%%





\bibitem{madgraph}
 F.~Maltoni and T.~Stelzer,
 ``MadEvent: Automatic event generation with MadGraph,''
 JHEP {\bf 0302}, 027 (2003)
 [arXiv:hep-ph/0208156].
 %%CITATION = JHEPA,0302,027;%%
\bibitem{bridge}
 P.~Meade and M.~Reece,
 ``BRIDGE: Branching ratio inquiry / decay generated events,''
 arXiv:hep-ph/0703031.
 %%CITATION = HEP-PH/0703031;%%
\bibitem{pythia}
 T.~Sjostrand, S.~Mrenna and P.~Skands,
 ``PYTHIA 6.4 physics and manual,''
 JHEP {\bf 0605}, 026 (2006)
 [arXiv:hep-ph/0603175].
 %%CITATION = JHEPA,0605,026;%%

 \bibitem{atlfast}
ALTFAST documentation (https://twiki.cern.ch/twiki/bin/view/Atlas/AtlfastDocumentation).


\bibitem{slep}
 M.~Fairbairn, A.~C.~Kraan, D.~A.~Milstead, T.~Sjostrand, P.~Skands and T.~Sloan,
  ``Stable massive particles at colliders,''
  Phys.\ Rept.\  {\bf 438}, 1 (2007)
  [arXiv:hep-ph/0611040].
 %%CITATION = PRPLC,438,1;%%


\bibitem{cms-tdr}
  G.~L.~Bayatian {\it et al.}  [CMS Collaboration],
 ``CMS technical design report, volume II: Physics performance,''
 J.\ Phys.\ G {\bf 34}, 995 (2007).
 %%CITATION = JPHGB,G34,995;%%
\bibitem{Allanach}
 B.~C.~Allanach, C.~M.~Harris, M.~A.~Parker, P.~Richardson and B.~R.~Webber,
 ``Detecting exotic heavy leptons at the Large Hadron Collider,''
 JHEP {\bf 0108}, 051 (2001)
 [arXiv:hep-ph/0108097];
 %%CITATION = JHEPA,0108,051;%%
 S.~Ambrosanio, B.~Mele, A.~Nisati, S.~Petrarca, G.~Polesello, A.~Rimoldi and G.~Salvini,
  ``SUSY long-lived massive particles: Detection and physics at the LHC,''
  arXiv:hep-ph/0012192;
   %%CITATION = HEP-PH/0012192;%%
 S.~Ambrosanio, B.~Mele, S.~Petrarca, G.~Polesello and A.~Rimoldi,
  ``Measuring the SUSY breaking scale at the LHC in the slepton NLSP  scenario of GMSB models,''
  JHEP {\bf 0101}, 014 (2001)
  [arXiv:hep-ph/0010081].
 %%CITATION = JHEPA,0101,014;%%

\bibitem{tulika}
Tulika Bose, private communication.

\bibitem{prospino}

W.~Beenakker, R.~Hopker and M.~Spira,
 ``PROSPINO: A program for the PROduction of Supersymmetric Particles In Next-to-leading Order QCD,''
 arXiv:hep-ph/9611232;
   %%CITATION = HEP-PH/9611232;%%
PROSPINO2: http://www.thphys.uni-heidelberg.de/~plehn/prospino/.


\bibitem{Mangano:2002ea}
  M.~L.~Mangano, M.~Moretti, F.~Piccinini, R.~Pittau and A.~D.~Polosa,
  ``ALPGEN, a generator for hard multiparton processes in hadronic
  collisions,''
  JHEP {\bf 0307}, 001 (2003)
  [arXiv:hep-ph/0206293].
  %%CITATION = JHEPA,0307,001;%%

\bibitem{diracphe}
S.~Y.~Choi, M.~Drees, A.~Freitas and P.~M.~Zerwas,
 ``Testing the Majorana Nature of Gluinos and Neutralinos,''
  Phys.\ Rev.\  D {\bf 78}, 095007 (2008)
  [arXiv:0808.2410 [hep-ph]];
    %%CITATION = PHRVA,D78,095007;%%
J.~Alwall, D.~Rainwater and T.~Plehn,
  ``Same-Sign Charginos and Majorana Neutralinos at the LHC,''
  Phys.\ Rev.\  D {\bf 76}, 055006 (2007)
  [arXiv:0706.0536 [hep-ph]].
    %%CITATION = PHRVA,D76,055006;%%


 %\cite{Chacko:2003tf}
\bibitem{Chacko:2003tf}
 Z.~Chacko and E.~Ponton,
``Bulk gauge fields in warped space and localized supersymmetry breaking,''
 JHEP {\bf 0311}, 024 (2003)
 [arXiv:hep-ph/0301171].
 %%CITATION = JHEPA,0311,024;%%

%\cite{Medina:2006hi}
\bibitem{Medina:2006hi}
 A.~D.~Medina and C.~E.~M.~Wagner,
 ``Soft leptogenesis in warped extra dimensions,''
 JHEP {\bf 0612}, 037 (2006)
 [arXiv:hep-ph/0609052].
 %%CITATION = JHEPA,0612,037;%%

  \bibitem{Kaplan:1999ac}
  D.~E.~Kaplan, G.~D.~Kribs and M.~Schmaltz,
 ``Supersymmetry breaking through transparent extra dimensions,''
  Phys.\ Rev.\  D {\bf 62}, 035010 (2000)
  [arXiv:hep-ph/9911293];
  %%CITATION = PHRVA,D62,035010;%%
  Z.~Chacko, M.~A.~Luty, A.~E.~Nelson and E.~Ponton,
 ``Gaugino mediated supersymmetry breaking,''
  JHEP {\bf 0001} (2000) 003
  [arXiv:hep-ph/9911323].
  %%CITATION = JHEPA,0001,003;%%

\bibitem{highscale}
 M.~Schmaltz and W.~Skiba,
  ``The superpartner spectrum of gaugino mediation,''
  Phys.\ Rev.\  D {\bf 62}, 095004 (2000)
  [arXiv:hep-ph/0004210];
   %%CITATION = PHRVA,D62,095004;%%
    M.~Schmaltz and W.~Skiba,
  ``Minimal gaugino mediation,''
  Phys.\ Rev.\  D {\bf 62}, 095005 (2000)
  [arXiv:hep-ph/0001172];
    %%CITATION = PHRVA,D62,095005;%%
 H.~Baer, A.~Belyaev, T.~Krupovnickas and X.~Tata,
  ``The reach of the Fermilab Tevatron and CERN LHC for gaugino mediated  SUSY
  breaking models,''
  Phys.\ Rev.\  D {\bf 65} (2002) 075024
  [arXiv:hep-ph/0110270].
  %%CITATION = PHRVA,D65,075024;%%
}
\end{thebibliography}
\end{document}